\documentclass[10pt,journal,compsoc,letterpaper]{IEEEtran}

\usepackage{cite}
\usepackage{graphicx}
\usepackage{textcomp}
\usepackage{xcolor}
\usepackage{hyperref}
\hypersetup{
    hidelinks
}

\makeatletter
\newcommand{\TPDSpubnote}[1]{%
  \begingroup
  \renewcommand{\thefootnote}{}%
  \let\@IEEEcompsocmakefnmark\@empty
  \footnotetext{#1}%
  \endgroup
}
\makeatother
\usepackage{booktabs}
\usepackage{tabularx}
\usepackage{multirow}

\usepackage{bm}
\usepackage{algorithm}
\usepackage[noend]{algpseudocode}   

\usepackage[caption=false,font=footnotesize,labelfont=sf,textfont=sf]{subfig}

\usepackage{amsmath,amssymb,amsfonts}

\usepackage{enumitem}
\setlist[itemize]{
  topsep=1.5pt,
  partopsep=0pt,
  parsep=0pt,
  itemsep=0pt,
  leftmargin=*
}

\usepackage[most]{tcolorbox}

\definecolor{greyback}{gray}{0.96}
\tcbset{
  insightbox/.style={
    enhanced,
    boxrule=0pt,
    colback=greyback,
    colframe=greyback,
    borderline west={2pt}{0pt}{black},
      left=8pt,
      right=8pt,
      top=6pt,
      bottom=6pt,
      before skip=10pt,
      after skip=10pt,
      fontupper=\normalsize
  }
}

\newtcbox{\newcode}{
  on line,
  box align=base,
  arc=3pt,
  colback=gray!15,
  colframe=gray!75!black,
  boxrule=0.5pt,
  left=2pt,
  right=2pt,
  top=2pt,
  bottom=2pt,
  boxsep=1pt
}

\usepackage{soul}  
\usepackage{url}
\newcommand{\sysname}{\textsf{BandPilot}}

\begin{document}
\bstctlcite{IEEEexample:BSTcontrol}

\title{\sysname{}: Toward Performance- and Contention-Aware GPU Dispatching in AI Clusters}

\author{Kunming~Zhang,
        Hanlong~Liao,
        Junyu~Xue,
        Deke~Guo,
        and~Guoming~Tang%
\thanks{
Kunming Zhang and Guoming Tang are with the Data Science and Analytics Thrust,
The Hong Kong University of Science and Technology (Guangzhou),
Guangzhou, China (e-mail: kzhang519@connect.hkust-gz.edu.cn; guomingtang@hkust-gz.edu.cn).}%
\thanks{Hanlong Liao is with the College of System Engineering,
National University of Defense Technology, Changsha, China
(e-mail: hanlongliao@nudt.edu.cn).}%
\thanks{Junyu Xue is with the Department of Computer Science and Engineering,
Southern University of Science and Technology, Shenzhen, China
(e-mail: junyuxue@outlook.com).}
\thanks{Deke Guo is with the School of Computer Science and Engineering,
Sun Yat-sen University, Guangzhou, China
(e-mail: guodk@mail.sysu.edu.cn). Corresponding author: Guoming Tang.}%
}



\maketitle

\TPDSpubnote{%
Published in \textit{IEEE Transactions on Parallel and Distributed Systems} (TPDS), Vol.~37, No.~10, 2026,
Doi: 10.1109/TPDS.2026.3716225.
}

\begin{abstract}
Modern multi-tenant AI clusters are increasingly communication-bound, driven by high-volume and multi-round GPU-to-GPU collective communication. Consequently, the GPU dispatcher’s choice of a physical GPU subset for each tenant largely determines the job’s effective collective bandwidth and thus its performance ceiling. Existing dispatchers predominantly rely on static, topology-aware heuristics that prioritize GPU resource compactness, assuming that minimizing physical distance maximizes communication bandwidth. 
 However, we reveal that this assumption often fails due to complex system-level bottlenecks, such as non-linear NIC saturation and inter-node link heterogeneity. 
This paper presents \sysname{}, a performance- and contention-aware GPU dispatching primitive that optimizes effective collective bandwidth for multi-tenant AI clusters. Specifically,  \sysname{} learns a data-efficient bandwidth model from sparse NCCL measurements via a hierarchical design.
Guided by the model, \sysname{}  uses an equilibrium-driven heuristic as a fast front end, and invokes a pruned elimination search  when a controller predicts that further refinement is worthwhile.
To account for multi-tenant interference, \sysname{} virtually merges a candidate allocation with co-located cross-host jobs to conservatively estimate shared bottleneck capacity and predict contention-degraded bandwidth. 
Across a 32-GPU H100 cluster and heterogeneous simulations, \sysname{} achieves 90 \text{--} 97\% bandwidth efficiency relative to the best-found reference, improving average efficiency by 20 \text{--} 30\% over topology-compactness heuristics. 

\end{abstract}



\section{Introduction}

\begin{figure}[t]
  \centering
  \includegraphics[width=0.85\linewidth]{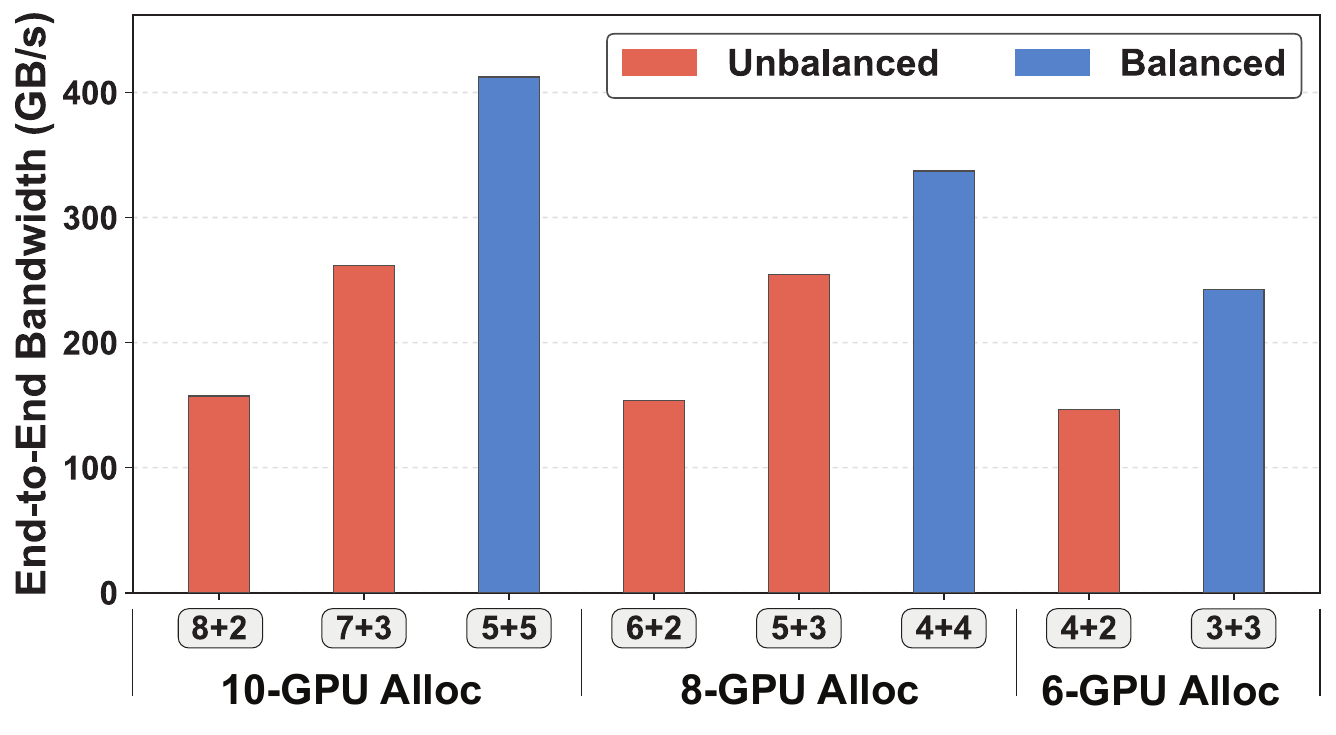}
  \caption{Effective NCCL (NVIDIA Collective Communications Library) \texttt{all-reduce} bandwidth on a real-world H100 cluster. For an 8-GPU request, a balanced \newcode{4+4} allocation across two nodes delivers over $2.2\times$ the bandwidth of an unbalanced \newcode{6+2} allocation favored by existing topology-aware dispatchers.}
  \label{fig:balance and unbalance}
\end{figure}

Driven by communication-intensive workloads such as large language model (LLM) training and inference~\cite{smith2022usingdeepspeedmegatrontrain,jiang2024megascale}, large-scale multi-accelerator AI clusters have become the cornerstone of modern cloud infrastructure. In these environments, especially for communication-intensive workloads, the effective communication bandwidth among the allocated GPUs is a key upstream factor for end-to-end performance~\cite{hidayetoglu2024commbench,siefert2023latency}. As cloud providers increasingly expose GPUs as a service, a tenant typically requests a bundle of $k$ GPUs, and a resource manager must dispatch a concrete subset from a shared pool. This seemingly simple \emph{GPU dispatching} decision fixes the physical communication topology for the job and thus sets the upper bound for all subsequent communication optimizations. An imprudent initial allocation can irrevocably cripple performance, regardless of later improvements in collective communication algorithms or congestion control. Congestion-aware fabrics can mitigate runtime contention after traffic is admitted, but they do not eliminate the \emph{structural} bottleneck created by a poor GPU choice at dispatch time.

Despite this central role, current GPU dispatchers still rely on static, locality-driven heuristics. The long-standing \textit{\textbf{proximity-based rule}}, also the \textit{default} dispatcher in Slurm and many production systems~\cite{nvtags2023,martinasso2018rmslurm}, greedily co-locates GPUs within the same server or socket to ensure short and uniform communication paths. While adequate for single workstations with homogeneous PCIe links, this strategy falters in multi-rack, multi-tier fabrics where link capacities can vary by an order of magnitude. To better adapt to such fabrics, Slurm further provides a \textit{\textbf{topology-aware strategy}}~\cite{soner2014topologically}: based on a manually created topology file that assigns weights to different link types (e.g., NVLink, PCIe switch, inter-node), the dispatcher greedily selects the most ``compact'' GPU set that fulfills the request, prioritizing GPUs that span the fewest nodes and switches to maximize locality and minimize resource fragmentation. Unfortunately, both strategies fundamentally \emph{conflate resource compactness with communication performance}.

Our empirical analysis on a real-world 32-GPU H100 cluster reveals that this compactness heuristic can lead to profoundly suboptimal outcomes. As shown in Fig.~\ref{fig:balance and unbalance}, when dispatching 8 GPUs across two nodes, each with 6 idle GPUs, the topology-aware dispatcher yields an unbalanced \newcode{6+2} allocation (to maximize locality). A more balanced \newcode{4+4} allocation, while less compact, delivers more than double the effective NCCL bandwidth (337.17~GB/s vs.\ 153.44~GB/s).  The root cause is a system-level bottleneck that static topology rules fail to capture: in modern multi-tier fabrics, the balance of load across Network Interface Cards (NICs) and inter-node links can dominate compactness as the primary determinant of end-to-end bandwidth, especially in the presence of heterogeneity and NUMA effects~\cite{EvaluatingModernGPUInterconnect}.

In isolation, these examples might suggest that one only needs a better notion of ``balanced'' allocations in an idle cluster. However, production AI clusters are inherently \emph{multi-tenant}. When a new request arrives, a substantial fraction of GPUs is already leased to other tenants; these tenants generate background traffic on bandwidth-critical NVLink, PCIe, and inter-node fabrics, reducing the residual bandwidth available to any new allocation. Consequently, the effective NCCL bandwidth of a candidate GPU subset is also determined by the prevailing contention on shared links and GPU devices across the cluster. Conceptually, the communication performance of an allocation can be viewed as a black-box outcome of the chosen GPUs and the current cluster state, which we later formalize in Sec.~\ref{sec:problem-formulation}.

An ideal dispatching strategy would thus be both \emph{performance-aware} and \emph{contention-aware}. In principle, it would possess prior knowledge of the communication performance attainable by every feasible allocation under the current cluster state, so that it can always select a near-optimal GPU subset for each request. Obtaining such knowledge directly is practically impossible. Communication performance emerges from complex interactions between hardware topology, NCCL's internal routing and algorithm choices, and the time-varying background load produced by opaque tenant workloads. The only reliable way to obtain ground-truth behavior is to execute real collective operations on the cluster. At the same time, the search space of candidate allocations grows combinatorially as $\binom{|A|}{k}$ for an available set $A$ and request size $k$, rendering exhaustive measurement or brute-force search intractable for real-time dispatching in realistic clusters. Any practical solution therefore requires a data-driven surrogate model that plays the role of this oracle while satisfying three stringent system requirements: \emph{(i) high predictive accuracy} to reliably guide allocation decisions, \emph{(ii) data efficiency} to avoid prohibitively expensive cluster-wide benchmarking, and \emph{(iii) architectural scalability} to handle heterogeneous, dynamic clusters with routine infrastructure changes without frequent retraining.

To this end, we propose and design \sysname{} and make the following contributions:
\begin{itemize}[leftmargin=*]
  \item \textbf{Problem formulation.}
  We formulate GPU dispatching as a contention-aware optimization problem over effective NCCL bandwidth, while treating jobs as black boxes without requiring model architectures or runtime traces. To the best of our knowledge, \sysname{} is the first to treat selecting a physical GPU subset as a first-class, performance- and contention-aware dispatching primitive for GPU rental in multi-tenant AI clusters. 

  \item \textbf{Performance- and contention-aware model.}
  \sysname{} introduces a hierarchical Transformer surrogate that predicts the standalone bandwidth of arbitrary GPU subsets from sparse NCCL measurements. On top of this surrogate, a contention-aware predictor virtually merges a candidate allocation with co-located jobs to conservatively estimate residual bottleneck capacity on shared NICs and inter-node fabrics.

  \item \textbf{Practical GPU dispatching system.}
  \sysname{} builds a bandwidth- and contention-aware GPU dispatching system that replaces proximity-based and topology-aware heuristics with a data-driven core. Guided by the predictor, a fast hybrid search couples an equilibrium-driven constructor, a runtime gate controller, and a pruned tree search to efficiently navigate the combinatorial allocation space.  \sysname{} identifies near-optimal allocations that can be plugged into existing schedulers as a control-plane primitive.

  \item \textbf{Implementation and evaluation.}
  We implement \sysname{} on a 32-GPU H100 cluster and in heterogeneous, trace-driven simulations.
  Across dynamic multi-tenant traffic profiles, \sysname{} consistently achieves \(90 \text{--} 97\%\) of the best-found bandwidth and improves average bandwidth efficiency by \(20 \text{--} 30\%\) over topology-compactness dispatchers. All the source code and measurement data are released at \url{https://github.com/SusCom-Lab/BandPilot}
\end{itemize}



\section{Background and Motivation}
\label{sec:background}

\subsection{The Role of GPU Dispatcher in AI Clusters}
\label{sec:Role of GPU Dispatcher}

The advent of large-scale AI and LLMs has cemented AI clusters, especially those comprising hundreds to thousands of GPUs, as the cornerstone of modern cloud infrastructure.
These clusters are evolving from monolithic, homogeneous systems into large-scale, heterogeneous environments~\cite{jiang2020unified}.
It is now commonplace to find different GPU generations and diverse interconnects coexisting within a single pool.
This architectural complexity, combined with the non-linear behavior of modern communication libraries, makes it profoundly challenging to predict performance using simple topology-based heuristics.

At the same time, the deployment model of such clusters is increasingly \emph{multi-tenant} and \emph{GPU-as-a-Service}~\cite{peng2018optimus,choi2022serving}.
A tenant typically requests a bundle of $k$ GPUs, while the provider exposes only a shared pool of partially occupied GPUs. 
The internal details of the tenant's workload, including model architecture, communication pattern, and runtime, are often opaque to the infrastructure. 
Within this setting, the \emph{GPU dispatcher} is the system entity responsible for selecting a concrete physical subset of GPUs from the currently available set.
Its decision is made per request, based on cluster-level information such as hardware topology, GPU availability, and optionally coarse-grained traffic statistics.
Once a subset is chosen, the communication topology for that job is fixed, and the resulting effective bandwidth sets the \emph{performance upper bound} that any downstream scheduler or communication optimization can hope to achieve. 
Consequently, an imprudent initial dispatch can irrevocably cripple an application's communication performance, a deficit that subsequent software-level optimizations cannot fully recover. 

Notably, as the dispatching process does not involve any job-level semantics, the GPU dispatcher in our context is not a replacement for existing job schedulers or communication libraries. Instead, it is a control-plane primitive that can be plugged into them.

\subsection{Topology-Compactness Heuristic and Pitfalls}
\label{sec:The Pitfalls of Compactness-driven Paradigm}

For GPU dispatching in AI clusters, state-of-the-art (SOTA) resource managers such as Slurm have adopted a \emph{topology-aware strategy}~\cite{martinasso2018rmslurm}.
The design philosophy of this strategy is to prioritize \emph{compactness} in order to maximize data locality and minimize resource fragmentation.
Given a request for $k$ GPUs, the dispatcher employs a greedy algorithm to find the most compact GPU allocation, favoring sets that span the fewest possible nodes and switches. The core assumption underlying this paradigm is that resource compactness is a reliable proxy for high communication performance.

Our empirical analysis on a real-world H100 GPU cluster reveals that this pursuit of compactness often leads to profoundly suboptimal outcomes, even when the cluster is otherwise idle.
As shown in Fig.~\ref{fig:balance and unbalance}, when dispatching 10 GPUs across two nodes (each with 8 idle GPUs), a topology-aware dispatcher selects a compact \newcode{8+2} allocation.
In contrast, a less compact but more balanced \newcode{5+5} allocation delivers more than double the effective communication bandwidth (412.49~GB/s vs.\ 157.30~GB/s).
Such a performance difference has substantial real-world impact. 
In our LLaMA-2 1B training experiment on NVIDIA A6000 GPUs, increasing the effective communication bandwidth from 5~GB/s to 22.5~GB/s reduces the per-iteration time to 66\% of the low-bandwidth case.


The root cause of this performance disparity lies in a system-level bottleneck that static, topology-based heuristics fail to capture.
In the \newcode{8+2} scenario, the collective communication operations saturate the NIC of the node hosting eight GPUs.
This single NIC becomes the bottleneck for all inter-node traffic, throttling the entire collective operation and dragging down the end-to-end bandwidth. This demonstrates a non-linear effect: the \emph{balance} of GPU allocation across nodes and NICs can be a more dominant performance factor than topological compactness. 

\begin{figure}[t]
  \centering
  \includegraphics[width=0.80\linewidth]{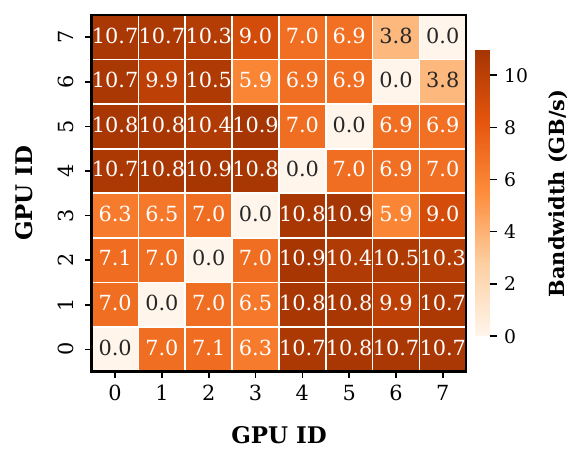}
  \caption{Peer-to-peer bandwidth measurements on 8 NVIDIA RTX 4090 GPUs. The P2P bandwidth between proximal GPUs (e.g., GPU~0 and GPU~1) can be lower than that between more remote pairs (e.g., GPU~0 and GPU~7), illustrating anti-locality effects.}  
  \label{fig:BG:4090_bw_heatmap}

\end{figure}

More broadly, as illustrated in Fig.~\ref{fig:BG:4090_bw_heatmap}, modern interconnect performance can be counter-intuitive due to architectural effects, including anti-locality behaviors where a more remote path is unexpectedly faster than local~\cite{EvaluatingModernGPUInterconnect}.

\begin{tcolorbox}[insightbox]
\textbf{Remark 1.}
{The load-balancing bottleneck exemplifies the complex and often counter-intuitive performance landscape of interconnects in modern AI clusters.
Existing compactness-driven dispatching strategies are fundamentally unreliable as surrogates for communication performance, and they provide no principled way to incorporate dynamic contention.}
\end{tcolorbox}

\subsection{Towards Performance- and Contention-Aware Dispatching}
\label{sec:challenge}

The deficiencies of static, compactness-driven heuristics highlight the need for a performance-aware approach to GPU dispatching.
In modern multi-tenant clusters, however, the effective communication performance of a candidate allocation depends not only on which GPUs are selected, but also on the \emph{current contention state} of the cluster.
When a new request arrives, a substantial fraction of GPUs is typically already leased to other tenants, whose collective communication generates background traffic on NVLink, PCIe, and inter-node links.
Static rules based on proximity or manually tuned link weights offer no principled way to incorporate such dynamic effects.

An ideal dispatcher would, in principle, know the resulting collective communication performance of every feasible GPU subset under the current cluster state, including hardware heterogeneity and ongoing traffic.
Armed with this knowledge, it could always select a near-optimal allocation for each incoming request. To acquire this knowledge, however, we face major challenges that render traditional methods intractable.

It is computationally prohibitive to exhaustively measure the end-to-end bandwidth of every GPU subset, especially in large-scale clusters.
This infeasibility necessitates a shift towards a practical methodology based on predictive modeling. The design and integration of such a model into a dispatcher must overcome three core system challenges.
\begin{itemize}
    \item \textbf{High predictive accuracy.}
    The predictive model acts as a stand-in for ground-truth communication performance.
    It is essential to be highly accurate as minor prediction errors can mislead the dispatcher, resulting in the selection of suboptimal GPU subsets. 
    \item \textbf{Data efficiency.} Sophisticated predictive models often require vast amounts of training data. However, collecting these data through exhaustive cluster benchmarking is itself intractable. A practical solution must therefore be data-efficient, capable of generalizing from a sparse set of bandwidth measurements.
    \item \textbf{Architectural scalability.} Production AI clusters are not static, undergoing hardware changes through upgrades, maintenance and failures. A viable model must be inherently scalable and elastic, be capable of processing variable-sized inputs and remain effective under regular changes.
\end{itemize}

\begin{tcolorbox}[insightbox]
\textbf{Remark 2.}
{To tackle these challenges, we argue for a new paradigm of \emph{performance- and contention-aware} GPU dispatching.
Such a system must be built upon a scalable, accurate, and data-efficient predictive model that can estimate the communication performance of candidate GPU subsets under realistic cluster states, including multi-tenant sharing.
Guided by this predictor, the dispatcher further requires an efficient search algorithm to navigate the vast combinatorial space and identify the near-optimal allocation at low overhead.}
\end{tcolorbox}

\section{Problem Formulation}
\label{sec:problem-formulation}

This section formalizes the contention-aware GPU dispatching problem. We first describe the system model and scope, then define the bandwidth-based optimization objective and evaluation metrics, and finally highlight the core difficulties and our high-level ideas.

\subsection{System Model}

\subsubsection{Setting and Scope}
We consider a general scenario where an AI cluster provider  offers bare-metal GPUs as a rental service. The provider exposes a \emph{GPU allocation primitive}: when a tenant issues a request, the system must choose a subset of GPUs for the tenant.
Unlike a job-level scheduler, the provider does not know the tenant's workload type and does not know the job duration.
Consequently, the provider cannot optimize the application-level step time directly, but can instead shape the \emph{maximum communication bandwidth} that the allocated GPUs can sustain.

\subsubsection{Cluster and request model}
We model the cluster as a finite set of GPUs
\[
\mathcal{G} = \{g_1, g_2, \dots, g_N\}.
\]
At any time, a subset of GPUs is currently idle and available for new requests; we denote this set by $\mathcal{A} \subseteq \mathcal{G}$.
A tenant request is abstracted as a demand for $k$ GPUs, where $k \leq |\mathcal{A}|$. We do \emph{not} assume any additional job-level information (e.g., model architecture or batch size).

A \emph{candidate allocation} is a size-$k$ subset of available GPUs, $S \subseteq \mathcal{A}$ with $|S|=k$.
The search space for the dispatcher is the set of all such allocations:
\begin{equation}
\mathcal{C}(\mathcal{A}, k)
= \{ S \subseteq \mathcal{A} : |S| = k \},
\end{equation}
whose cardinality $|\mathcal{C}(\mathcal{A}, k)| = \binom{|\mathcal{A}|}{k}$ grows combinatorially with both the number of available GPUs and the request size.
\sysname{} operates in a \emph{per-request}  regime: for each arriving request, it chooses one allocation $S \in \mathcal{C}(\mathcal{A}, k)$ under the \emph{current} cluster state.

\subsubsection{Background traffic and traffic profile}
At the time a new request arrives, some GPUs may already be leased to other tenants.
Let $\mathcal{J}$ denote the set of active jobs executed by previous tenants, and let $S_j \subseteq \mathcal{G}$ denote the GPU subset assigned to job $j \in \mathcal{J}$.
These active jobs generate background communication traffic on bandwidth-critical links, which reduces the bandwidth attainable by any newly arriving tenant.

Modeling the detailed packet-level behavior of all active jobs is not desirable in a cloud control-plane. We therefore summarize the current background state by a \emph{traffic profile} 
\[
\tau \in \mathcal{T},
\]
where $\tau$ denotes the contention-relevant background state exposed to the dispatcher. In this section, $\tau$ is treated as an abstract parameter that captures the occupancy and coarse communication demand of active jobs, rather than a per-flow or per-link traffic matrix. \sysname{}'s concrete instantiation of $\tau$ is introduced in Sec.~\ref{sec:contention-predictor}.  In the remainder of this section, we treat $\tau$ as a given parameter describing the current contention state and use it to define the standalone and contention-aware bandwidth functions for dispatch.
  
\subsubsection{Bandwidth functions}
We focus on the bandwidth of a fixed NCCL collective as the dispatch-level performance proxy.
For any subset $S \subseteq \mathcal{G}$, we define the \emph{standalone} bandwidth
\begin{equation}
B(S) \in \mathbb{R}_{\geq 0}
\end{equation}
as the effective bandwidth achieved when $S$ runs in isolation, with no competing background traffic.
Operationally, $B(S)$ is obtained by running an \emph{nccl-tests}~\cite{nvidia_nccl-tests} benchmark on $S$ under an idle cluster.

Under an arbitrary traffic profile $\tau \in \mathcal{T}$, the same allocation $S$ no longer attains $B(S)$ because active jobs consume part of the shared network capacity.
We model the \emph{effective} bandwidth delivered to a new job placed on $S$ as
\begin{equation}
B(S, \tau) \in \mathbb{R}_{\geq 0},
\end{equation}
which we interpret as the maximum collective bandwidth that the cluster can offer to the new tenant in the presence of background traffic summarized by $\tau$.
By construction,
\begin{equation}
0 \leq B(S, \tau) \leq B(S), \quad \forall S \subseteq \mathcal{G},\ \forall \tau \in \mathcal{T},
\end{equation}
and $B(S, \tau_{\mathrm{idle}}) = B(S)$ for a suitable ``idle'' profile $\tau_{\mathrm{idle}}$ representing negligible contention.

\subsection{Optimization Objective}

Given a request for $k$ GPUs, the current available set $\mathcal{A}$, and a traffic profile $\tau$, the dispatcher must select an allocation $S \in \mathcal{C}(\mathcal{A}, k)$.
\sysname{} optimizes for communication performance by treating the $B(S, \tau)$ as the objective, rather than application-level step time.
Formally, the \emph{contention-aware GPU subset selection} problem is
\begin{equation}
\label{eq:optimal_problem}
 S^*(\mathcal{A}, k, \tau)
= \arg\max_{S \in \mathcal{C}(\mathcal{A}, k)} B(S, \tau).
\end{equation}

Eq.~(\ref{eq:optimal_problem}) defines a per-request optimization problem: each arriving request is handled independently under the current cluster snapshot $(\mathcal{A}, \tau)$.
We do not attempt to jointly optimize over unknown future arrivals or to minimize long-term resource fragmentation.

For a given request and traffic profile we denote the corresponding optimal bandwidth by
\begin{equation}
B^*(\mathcal{A}, k, \tau)
= \max_{S \in \mathcal{C}(\mathcal{A}, k)} B(S, \tau).
\end{equation}

To normalize across different scenarios for evaluation, we further define the \emph{GPU bandwidth efficiency (GBE)} of an allocation $S$ as
\begin{equation}
\label{algo:metric}
\text{GBE}(S; \mathcal{A}, k, \tau)
= \frac{B(S, \tau)}{B^*(\mathcal{A}, k, \tau)}.
\end{equation}
In particular, we denote $S_{\text{sol}}$ as the allocation produced by a given dispatcher, then $\text{GBE}(S_{\text{sol}}; \mathcal{A}, k, \tau)$ lies in $[0,1]$ and approaches $1$  when the dispatcher finds a near-optimal allocation under the current background traffic.


\subsection{Difficulties and High-level Ideas}

Solving the contention-aware optimization problem in Eq.~(\ref{eq:optimal_problem}) directly in a production cluster is intractable due to two fundamental difficulties.

\begin{itemize}
    \item \textbf{Black-box, contention-dependent objective.}
    It is hard to model the effective bandwidth $B(S, \tau)$ by a closed-form analytical expression.
    It depends on the cluster's topology, NCCL's internal routing and algorithm choices, and the instantaneous traffic profile $\tau$ induced by opaque tenant workloads.
    Its value for a given allocation $S$ and profile $\tau$ can only be obtained through empirical measurement on the cluster under the current background load, which is too expensive to perform exhaustively.

    \item \textbf{Combinatorial search space.}
    Even for a fixed request size $k$ and available set $\mathcal{A}$, the search space $\mathcal{C}(\mathcal{A}, k)$ has size $\binom{|\mathcal{A}|}{k}$.
    Enumerating and empirically evaluating all allocations is infeasible for realistic clusters with tens or hundreds of available GPUs.
\end{itemize}

 Existing dispatchers rely on topology-compactness heuristics that ignore $\tau$ and thus are poor proxies for the true contention-dependent objective $B(S, \tau)$.
\sysname{} addresses these difficulties with two complementary ideas.

\subsubsection{Bandwidth surrogate model}
We construct a data-driven surrogate $\hat{B}(S, \tau)$ that predicts the effective bandwidth for any allocation and traffic profile.
The surrogate is trained from a sparse set of NCCL measurements and is designed to be efficient to evaluate, enabling contention-aware reasoning over many candidate allocations without exhaustively benchmarking the cluster.

\subsubsection{Efficient search algorithm}
Guided by the surrogate $\hat{B}(S, \tau)$, \sysname{} employs a specialized search algorithm that explores the combinatorial space $\mathcal{C}(\mathcal{A}, k)$ while keeping the number of surrogate evaluations small. The resulting allocation $S_{\text{sol}}$ approximately maximizes $B(S, \tau)$ and thus achieves high GBE under contention, providing a practical solution to the contention-aware GPU subset selection problem.

\section{System Design}

To solve the optimal GPU dispatching problem defined in Sec.~\ref{sec:problem-formulation}, we propose and design \sysname{}, a data-driven GPU dispatching system with a performance-aware core. This section details the architecture of \sysname{}, which is engineered to address the challenges of prediction accuracy, data and search efficiency, and architectural scalability in a production environment.

\subsection{Architectural Overview}
\label{sec:design:overview}

\begin{figure}[t!]
  \centering
  \includegraphics[width=\linewidth]{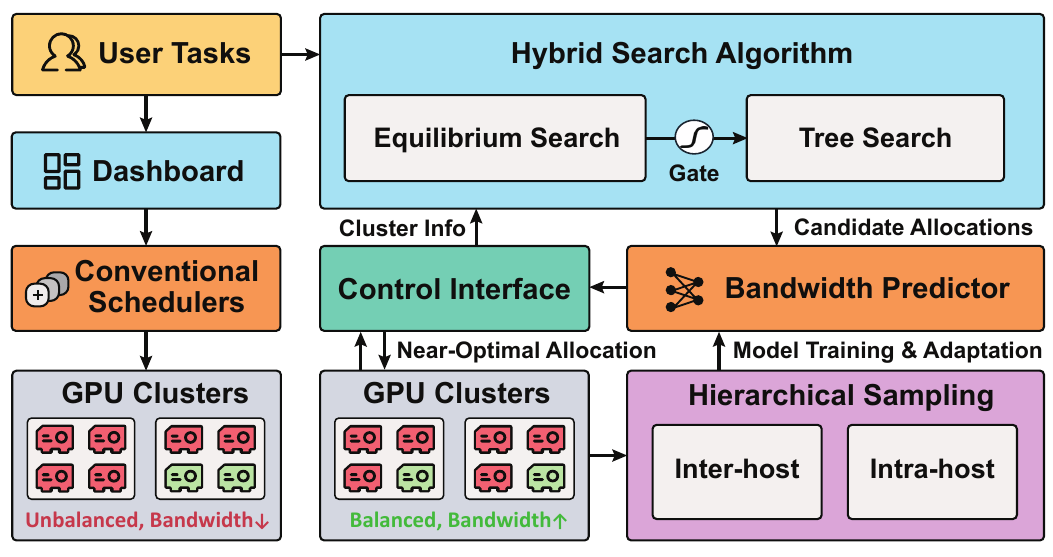}
  \caption{Architectural overview of the \sysname{} system. The left path depicts a conventional scheduler, whose topology-driven decisions can result in unbalanced allocations and suboptimal bandwidth. The right main workflow illustrates \sysname{}'s data-driven approach, which identifies near-optimal, balanced allocations. }
  \label{fig:overview}
\end{figure}

As shown in Fig.~\ref{fig:overview}, \sysname{} separates per-request dispatch from offline profiling and background model maintenance. At dispatch time, the Control Interface constructs the current cluster snapshot, and the Bandwidth Predictor and Hybrid Search Algorithm jointly select a near-optimal GPU subset for the incoming request. Outside the critical path, Hierarchical Sampling performs one-time bring-up profiling for each host type, while Model Training and Adaptation updates the inter-host surrogate with newly collected measurements at low frequency.

\subsubsection{Online Dispatch Path}
When a request for $k$ GPUs arrives, \sysname{} serves it through a lightweight online dispatch path. 

\textbf{Control Interface.}
The request first enters the Control Interface, which maintains the dispatch-time state of the cluster, including the set of available GPUs $\mathcal{A}$ and the traffic profile $\tau$. It exposes this state to the dispatcher core as the input context for allocation selection.

\textbf{Dispatcher core.}
Given $(\mathcal{A}, k, \tau)$, the dispatcher core evaluates the request under the current cluster snapshot. The Bandwidth Predictor estimates the $\hat{B}(S,\tau)$ for candidate $S$, and the Hybrid Search Algorithm efficiently explores the combinatorial search space $\mathcal{C}(\mathcal{A}, k)$ and identifies a near-optimal allocation $S_{\mathrm{sol}}$.

\textbf{Deployment.}
The $S_{\mathrm{sol}}$ is returned to the Control Interface, which provisions the corresponding physical GPUs for the incoming request.

\subsubsection{Model Initialization and Adaptation Path}
The predictor is supported by a separate initialization and adaptation path to keep it accurate. 

\textbf{Initialization.}
During system bring-up, Hierarchical Sampling performs one-time offline profiling by exhaustively measuring intra-host bandwidths for each host type and sparsely sampling inter-host bandwidths across the cluster using \textit{nccl-tests}. These measurements initialize the hierarchical predictor and train the inter-host surrogate.

\textbf{Low-frequency adaptation.}
After deployment, \sysname{} can incorporate newly collected communication measurements through a low-frequency adaptation loop, allowing \sysname{} to preserve prediction accuracy under moderate drift and routine infrastructure changes.



\subsection{Bandwidth Surrogate Model}\label{sec:Lightweight Hierarchical Transformer}

To address the intractable problem of direct bandwidth measurement, we first introduce a surrogate model, $\hat{B}(S)$, designed to provide bandwidth information. In Sec.~\ref{sec:contention-predictor}, we will wrap it with a contention model to obtain $\hat{B}(S,\tau)$. As argued in Sec.~\ref{sec:challenge}, the design of this model is governed by three critical system requirements: accuracy, data efficiency and scalability.

We select the \textbf{Transformer} architecture as it natively satisfies our design requirements. Its core self-attention mechanism is designed to operate on sets of entities, making it inherently capable of processing variable-length sequences. This directly solves the input-size scalability problem: the model can accept any allocation $S$ of size $k$ from a pool $\mathcal{A}$ of any size, without any architectural changes. The self-attention mechanism computes a weighted representation for each GPU in the allocation by attending to all other GPUs in the same allocation. This allows the model to learn the complex, non-linear dependencies and interactions between all members of the subset $S$, which is essential for accurately predicting the collective bandwidth. 

\begin{figure}[t]
  \centering
  \includegraphics[width=\linewidth]{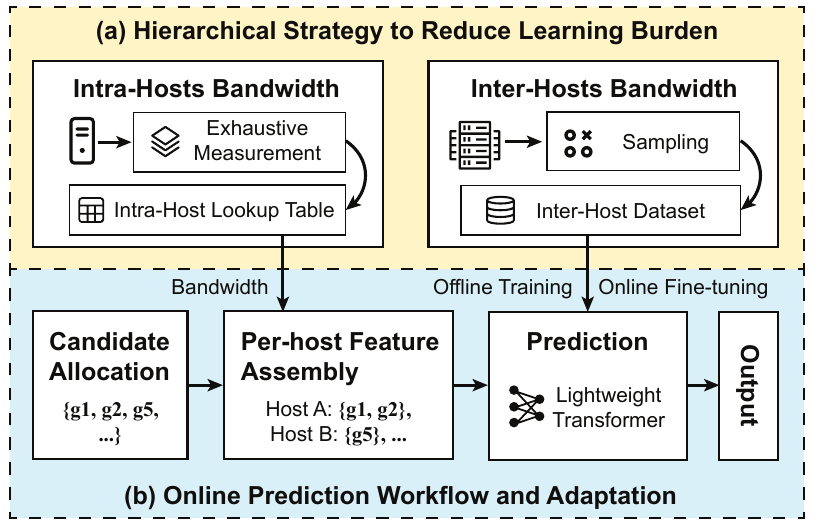}
  \caption{Hierarchical bandwidth prediction. Instead of a single monolithic model, we use measured lookups for intra-host bandwidth and a Transformer to predict the inter-host communication dynamics, significantly reducing the learning burden.}
  \label{fig:Transformer overview}
\end{figure}


\subsubsection{A Hierarchical Approach to Reduce Learning Burden}

A naive, end-to-end application of the Transformer, where the model predicts the final bandwidth from raw GPU identifiers, yielded poor performance in our initial experiments. This is because the model is burdened with learning the entire physical hierarchy from scratch, from the on-chip links to inter-node network dynamics. To overcome this, we introduce a hierarchical strategy that decomposes the prediction problem with a ``divide and conquer'' approach. We separate the problem into two distinct, more manageable stages, as illustrated in Fig.~\ref{fig:Transformer overview}.

\textbf{Stage-1: Deterministic intra-host bandwidth.} 
Communication within a single host is governed by a fixed and well-defined topology. The number of GPUs per host is typically small and fixed (e.g., 8). This makes it feasible to exhaustively measure the end-to-end collective bandwidth for all possible GPU combinations within a single host ($2^8 - 1 = 255$ combinations for an 8-GPU host). This is a one-time, offline process. The results are stored in a simple key-value dictionary for each host, mapping a specific GPU configuration tuple to its ground-truth bandwidth. This step replaces a complex learning problem with a fast, accurate lookup, providing a strong, data-driven foundation for the next stage. 

\textbf{Stage-2: Transformer for inter-host dynamics.}
With precise intra-host bandwidths available via lookup, the learning task of the Transformer is simplified. It is no longer required to predict the absolute bandwidth from scratch. Instead, its new objective is to learn the complex effects of \textit{inter-host communication} and how it interacts with the known intra-host performance.

The input to the Transformer is a sequence of feature vectors, one for each host that has at least one GPU in the allocation $S$. The feature vector for a given host $i$ contains: i) the pre-computed, intra-host bandwidth for the GPUs selected on that host from Stage-1 lookup, and ii) the number of GPUs selected on that host.

This hierarchical approach provides the model with highly informative features, allowing a lightweight Transformer to focus on modeling the inter-dependencies between nodes. This strategy has proven highly effective in Sec.~\ref{sec:ablation study}.

\subsubsection{Lightweight Architecture \&  Adaptation}
To ensure low latency and minimal storage footprint, our final model is deliberately lightweight. It consists of only 6 Transformer encoder layers with a hidden state dimension of 32, followed by a small 3-layer MLP prediction head.

To preserve accuracy under moderate drift, \sysname{} supports low-frequency updates using measurements collected after deployment. If the host-local topology is unchanged, the Stage-1 intra-host lookup tables remain valid. Routine events such as capacity expansion may slightly perturb the inter-host topology. In these cases, \sysname{} incrementally adapts Stage-2 rather than rerunning full offline retraining. Fundamental changes in fabric semantics, however, still require full offline retraining.


\subsection{Contention-aware Predictor}
\label{sec:contention-predictor}

The surrogate model above is trained on measurements collected on an idle cluster and thus approximates the $B(S)$ for $S \subseteq \mathcal{G}$. The optimization problem in Sec.~\ref{sec:problem-formulation}, however, is defined over the contention-dependent effective bandwidth $B(S,\tau)$ under the current background traffic profile $\tau$. To close this gap, \sysname{} augments the  $\hat B(S)$  with a contention model that maps the current cluster state to 
\begin{equation}
  \hat B(S,\tau) \approx B(S,\tau),
\end{equation}
and exposes $\hat B(S,\tau)$ to the search algorithm as the objective for ranking candidate allocations. It provides an approximation that captures the dominant interaction between the incoming tenant's communication \emph{demand} and the residual \emph{capacity} of the shared fabric. 


\subsubsection{State abstraction}

At dispatch time, the cluster hosts a set of active jobs $\mathcal{J}$. For each job $j \in \mathcal{J}$, \sysname{} records its allocated GPU subset $S_j \subseteq \mathcal{G}$
and a scalar bandwidth demand $d_j \ge 0$, which represents a coarse demand signal on the inter-node fabric. The collection $\{(S_j,d_j)\}_{j\in\mathcal{J}}$ is \sysname{}'s concrete instantiation of the traffic profile $\tau$ introduced in Sec.~\ref{sec:problem-formulation}. Here, $S_j$ comes directly from scheduler placement records, while $d_j$ is aggregated from cluster-level telemetry (e.g., DCGM counters) into a job-level demand signal rather than a packet-level measurement. 


Let $h(g)$ denote the host that contains GPU $g$, and for any allocation $S$ define the host set
\begin{equation}
  H(S) = \{ h(g) : g \in S \}.
\end{equation}
We call an allocation \emph{cross-host} if $|H(S)| \ge 2$. Since intra-host traffic stays within the local topology, \sysname{} only models contention among cross-host
allocations that share at least one host and hence compete for the same NIC or inter-node fabric.


Accordingly, for a candidate allocation $S$ under profile $\tau$, the predictor first extracts the subset of potentially interfering jobs:
\begin{equation*}
  \mathcal{J}_c(S,\tau) \;=\;
  \bigl\{ j \in \mathcal{J} :
    |H(S_j)| \ge 2 \;\text{ and }\; H(S_j) \cap H(S) \neq \emptyset
  \bigr\}.
\end{equation*}
Jobs whose host sets are disjoint from $H(S)$ are excluded, as they do not consume the bandwidth-critical inter-host resources relevant to $S$.

\subsubsection{Pairwise ``super allocations'' and bottleneck capacity}
For each $j \in \mathcal{J}_c(S,\tau)$, \sysname{} builds a synthetic \emph{super allocation} that aggregates the GPUs of the candidate and job $j$ on the hosts they share. Concretely, we construct an allocation that, for each host $n \in H(S) \cup H(S_j)$, contains as many GPUs as are used by the union $(S \cup S_j)$ on that host. This synthetic allocation captures the worst-case aggregate load placed by the pair $(S,S_j)$ on their shared NICs and upstream links, while abstracting away the exact intra-host topology. 
We then reuse the $\hat B(S)$ to estimate the bandwidth capacity that the shared inter-node fabric could provide to the two jobs together:
\begin{equation}
  \hat C_j(S,\tau) \;=\; \hat B\bigl(U_{S,j}\bigr),
\end{equation}
where $U_{S,j}$ denotes the super allocation for the pair $(S,S_j)$.%

The predictor derives a conservative bottleneck capacity for the candidate by taking the minimum over all relevant pairs:
\begin{equation}
  \hat C(S,\tau) \;=
    \min_{j \in \mathcal{J}_c(S,\tau)} \hat C_j(S,\tau)
\end{equation}
Intuitively, $\hat C(S,\tau)$ is an estimate of the collective bandwidth capacity that the inter-node fabric can simultaneously deliver to the candidate job and all cross-host jobs that share hosts with it.

\subsubsection{Demand aggregation and bandwidth sharing}
Let $d_S$ denote the demand of the candidate tenant, which \sysname{} sets to its standalone prediction,
\begin{equation}
  d_S \;=\; \hat B(S),
\end{equation}
and let $d_j$ be the stored demand of each contending job $j \in \mathcal{J}_c(S,\tau)$. The total demand on the shared bottleneck is
\begin{equation*}
  D(S,\tau) \;=\; d_S + \sum_{j \in \mathcal{J}_c(S,\tau)} d_j.
\end{equation*}
If the aggregate demand does not exceed the estimated bottleneck capacity, we have 
\begin{equation*}
  \hat B(S,\tau) \;=\; \hat B(S)
  \qquad\text{if } D(S,\tau) \le \hat C(S,\tau).
\end{equation*}

Otherwise, \sysname{} assumes that the candidate and its contending jobs share the bottleneck proportionally to their demands. The effective bandwidth offered to the candidate is then
\begin{equation*}
  \hat B(S,\tau) \;= \; \frac{d_S}{D(S,\tau)} \, \hat C(S,\tau)
  \qquad\text{if } D(S,\tau) > \hat C(S,\tau).
  \label{eq:contention-sharing}
\end{equation*}


This proportional-sharing rule is a dispatch-time approximation rather than a packet-level congestion model. Its purpose is to rank candidate allocations before placement, so \sysname{} emphasizes persistent bottleneck sharing instead of transient queueing dynamics.


\subsubsection{Interface to the dispatcher}
The contention-aware predictor is exposed to the dispatcher as a pure function that takes a candidate allocation $S$ and returns $\hat B(S,\tau)$ under the current cluster snapshot $(\mathcal{A},
  \tau)$. The hybrid search algorithm in Sec.~\ref{sec:Hybrid Search Algorithm} invokes this predictor repeatedly when exploring the combinatorial space $\mathcal{C}(\mathcal{A},k)$.

\subsection{Fast Hybrid Search Algorithm}\label{sec:Hybrid Search Algorithm}
Having established a fast and accurate bandwidth surrogate model, $\hat{B}(S)$, we now address the next challenge: efficiently searching the space $\mathcal{C}(\mathcal{A}, k)$ to find a high-quality allocation.

\subsubsection{The Hybrid Strategy}
To balance decision quality and runtime overhead, \sysname{} employs a hybrid search strategy with three components: 
(i) \textbf{Equilibrium-driven Heuristic Algorithm (EHA)},  a fast constructive method guided by the empirical insight that balanced allocations across hosts often yield superior bandwidth; 
(ii) \textbf{Pruned Tree Search (PTS)},  a more systematic top-down search that further refines difficult cases; and 
(iii) \textbf{K-Nearest Neighbors (KNN) gate}, a controller for the hybrid strategy, which predicts whether invoking PTS is worthwhile.

At runtime, \sysname{} executes EHA first. The KNN gate then uses EHA-derived confidence signals to decide whether additional PTS refinement is worthwhile to offset its overhead. If PTS is triggered, \sysname{} evaluates the EHA and PTS candidates under the same objective $\hat{B}(S,\tau)$ and returns the one with the higher score.

\subsubsection{Equilibrium-driven Heuristic Algorithm~(EHA)}
EHA is the fast constructive stage of \sysname{}. It is motivated by the observation that high-bandwidth allocations often correspond to load-balanced placements across active hosts. Instead of exploring the full search space, EHA builds a small set of high-potential candidates and scores them with $\hat{B}(S,\tau)$.

\textbf{Single-host prioritization.} If any host in $\mathcal{A}$ can satisfy a request for $k$ GPUs, EHA evaluates the best intra-host $k$-GPU subset on each such host and returns the best one.

\textbf{Balanced multi-host construction.} Otherwise, EHA finds the minimum number of active hosts needed for the request, forms a balanced base allocation, and generates a small canonical neighborhood of nearby feasible load splits. For larger clusters, this process is applied hierarchically across host groups and then refined within each group. The best candidate under $\hat{B}(S,\tau)$ is returned, together with compact candidate-set statistics used by the KNN gate.

\subsubsection{Pruned Tree Search (PTS)}
PTS serves as the systematic refinement stage of \sysname{}. While EHA rapidly constructs high-potential candidates, PTS explores the search space more deliberately through top-down elimination. It is motivated by the observation that collective communication performance is often dominated by the weakest link, so removing the most harmful resource can progressively improve the allocation.

\textbf{Coarse-to-fine elimination.} Starting from the full available set $S_{\mathrm{curr}}$, PTS iteratively removes one search unit at a time and keeps only the highest-scoring successor under $\hat{B}(S,\tau)$. The search first operates at coarse granularity, such as host-level or switch-local GPU groups, to discard clearly unfavorable regions quickly. Once further removal at the current granularity would violate feasibility, it switches to a finer granularity such as individual GPU and continues the refinement until $|S_{\mathrm{curr}}|=k$.
This design aggressively prunes the search tree into a single refinement path, making PTS substantially more systematic than EHA while still practical for online dispatch.



\subsubsection{K-Nearest Neighbors (KNN) gate}




Invoking PTS for every request would improve robustness but also add unnecessary online overhead. \sysname{} therefore places a KNN gate between EHA and PTS to predict when the extra refinement is worthwhile. We label PTS as worthwhile when
$(\hat{B}_{\mathrm{PTS}}-\hat{B}_{\mathrm{EHA}})/\hat{B}_{\mathrm{EHA+PTS}}\ge 0.05$. We adopt a KNN classifier because it is lightweight, non-parametric, and easy to update online.

The KNN gate uses features that summarize the difficulty and reliability of the EHA result. The feature vector contains the request size, EHA search evidence including the EHA node count and the host-count options explored by EHA, and placement-structure signals including the selected-active-node ratio, selected-node entropy, maximum selected-node share, half-$L_1$ distances from the selected distribution to the available and uniform distributions, and per-node selected-minus-available shares. 

At runtime, the gate estimates the risk of skipping PTS from the nearest labeled cases in the same cluster and traffic bucket. We use $k=5$ neighbors and require at least five effective support cases. The risk is the fraction of supported neighbors in which PTS was helpful. PTS is skipped only when this risk is at most $0.15$. Otherwise, including low-support or low-trust neighbor sets, the dispatcher falls back to PTS. 


The online policy is cold-started conservatively. Before activation,
\sysname{} runs both EHA and PTS and uses the observed outcomes as
shadow labels. The KNN gate is activated only after shadow execution observes
zero unsafe skips, an over-trigger rate below $50\%$ and no support-insufficient
cases. Here, an unsafe skip means that the shadow policy would skip PTS on a PTS-helpful case, an over-trigger means invoking PTS on a non-helpful case. To bound long-term lookup cost, each cluster-contention case base is capped at 1024 labeled cases with FIFO eviction.

\section{Evaluation}

In this section, we first detail the experimental setup, e.g., hardware platforms, and then assess the accuracy of bandwidth surrogate model and end-to-end dispatching performance. We also look into the system overhead and conduct ablation studies on the core model and algorithm designs.

\subsection{Experimental Setup}

\subsubsection{Hardware Platforms}\label{sec:Hardware_Platforms}
We conduct experiments on both a physical homogeneous cluster and a trace-driven, heterogeneous cluster simulation.
 Our experiments focus on clusters up to 32 GPUs, but the method itself does not assume a fixed and small scale.

\begin{table}[tb]
  \centering
  \small
  \caption{Hardware Platforms.}
  \label{tab:Hardware Platform}
  \resizebox{\columnwidth}{!}{
    \begin{tabularx}{\linewidth}{@{}ccc@{}}
      \toprule
      \textbf{Cluster Type} & \textbf{Composition} & \textbf{Short Name} \\
      \midrule
      Homogeneous   & 32$\times$ H100 & H100 \\
      \midrule
      Heterogeneous & 8 $\times$ \{4090, V100, A6000, A800\} & Het-4Mix \\
      \bottomrule
    \end{tabularx}
  }
\end{table}

\textbf{Homogeneous H100 cluster.} We deploy and evaluate our system on a physical cluster comprising 4 server nodes, each equipped with 8 NVIDIA H100 GPUs connected by NVSwitch. The nodes are interconnected via a 400,000 Mb/s ($\approx$ 50 GB/s) NVIDIA Quantum InfiniBand switch~\cite{nvidia_quantum2}, representing a high-bandwidth production environment. 

\textbf{Trace-based heterogeneous cluster simulations.} Due to the lack of access to a suitable heterogeneous physical cluster connected with switches, our evaluations were performed in a simulated environment constructed from bandwidth measurements on individual host types and the inter-host bandwidth data from the H100 cluster. We simulated a heterogeneous GPU cluster comprising RTX 4090, A6000, V100, and A800 devices, which differ in their native on-node interconnect topologies, ranging from PCIe-only to partially and fully connected NVLink. The simulation process involved:
\begin{itemize}
    \item \textbf{Intra-host measurement}. Measuring intra-host collective communication bandwidth for all combinations within each base host type independently using \textit{nccl-tests}.
    \item \textbf{Inter-host mapping}. We model the inter-host network fabric using the performance characteristics of
    our H100 cluster’s InfiniBand switch. To simulate a lower-grade network, we conservatively set the simulated switch bandwidth to
    one-fourth of the H100's bandwidth.
    \item \textbf{End-to-end bandwidth synthesis.} The effective
bandwidth for any given cross-node allocation is determined by the system’s bottleneck. It is calculated as the minimum of the pre-computed intra-host bandwidths of the involved hosts and the modeled inter-host link bandwidth.
\end{itemize}

Table~\ref{tab:Hardware Platform} details the configurations used in our evaluation. This simulation methodology allows us to create diverse cluster scenarios for our experiments. 

\textbf{Trace-based scale-out study.} For the large-scale overhead study in Sec.~5.4, we stress the online control path using a template-replicated Het-4Mix topology rather than a physical thousand-GPU deployment. We separate three evidence levels: measured dispatch latency on the 32-GPU H100 cluster, scaled search traces on 128--1024 GPUs, and synthesized latency bounds on 2048--4096 GPUs. The scaled traces execute the same EHA/KNN/PTS dispatch path as \sysname{}, while the synthesized bounds are derived from the scaled traces and profiled predictor costs.
Let $\Pi$ denote type-preserving permutations of the replicated topology. By construction, the scaled estimator $\tilde{B}$ is invariant to such type-preserving permutations:
\begin{equation}
\tilde{B}(S,\tau)=\tilde{B}(\pi(S),\pi(\tau)),\quad \forall \pi\in\Pi .
\end{equation}
Since \sysname{}'s online search uses only these type-preserving features and predicted scores, the dispatch procedure is equivariant in the same constructed topology
\begin{equation}
M(\pi(\mathcal{A}),k,\pi(\tau))=\pi(M(\mathcal{A},k,\tau)).
\end{equation}
Thus, the stress test evaluates how \sysname{}'s control-plane overhead grows with scheduling units, not general physical-bandwidth scaling under arbitrary thousand-GPU workloads.


\subsubsection{Bandwidth Measurement}
\sysname{} relies on measured bandwidth data for training the prediction model and populating intra-host dictionaries. 

We utilized the \textit{nccl-tests}~\cite{nvidia_nccl-tests} tool for this purpose. The \textit{nccl-tests} tool allows us to measure collective communication bandwidth for various patterns (e.g., \texttt{all-reduce}, \texttt{all-gather}) and data sizes. We profile multiple collective primitives and message sizes and observe that rankings across collectives are consistent, while smaller sizes (\texttt{2 MB}) are unstable and \texttt{16 MB} provides stable measurements with clear separation at acceptable runtime. Consequently, we standardize all measurements using \textit{nccl-tests} on \texttt{all-gather} with a \texttt{16 MB} payload.

\subsubsection{Traffic Profile}
\label{sec:evaluation traffic profile}

Following Sec.~\ref{sec:problem-formulation}, we model background network contention using a traffic profile $\tau \in \mathcal{T}$.  Because our evaluation is trace-driven rather than derived from deployed per-job telemetry, we use three representative profiles: $\tau_{\mathrm{idle}}$, $\tau_{\mathrm{mod}}$, and $\tau_{\mathrm{hvy}}$, to span no, moderate, and heavy contention, respectively. For each active job $j$ with placement $S_j$, we define its bandwidth demand $d_j$ as
\[
d_j =
\begin{cases}
0, & \tau = \tau_{\mathrm{idle}},\\
\rho_j \hat{B}(S_j), & \tau = \tau_{\mathrm{mod}},\\
\hat{B}(S_j), & \tau = \tau_{\mathrm{hvy}},
\end{cases}
\qquad
\rho_j \sim \mathcal{U}(0.25,\,0.75).
\]
Here, $\hat{B}(S_j)$ denotes the standalone inter-node bandwidth achievable by job $j$ under placement $S_j$.

\subsubsection{Benchmarks \& Metrics}

We compare \sysname{} against two groups of baselines. The first group contains the main dispatch policies:

\begin{itemize}
    \item \textbf{Proximity-based (\textit{Default}):} A heuristic that greedily packs GPUs within the same NUMA domain or CPU affinity group, without considering interconnect or switch structure.
    \item \textbf{Topology-aware (\textit{Topo}):} A SOTA heuristic that uses a static weighted topology graph to select the most compact GPU set and maximize interconnect locality.
    \item \textbf{CASSINI-inspired dispatch baseline (\textit{CasCore}):} A dispatch-level adaptation of CASSINI's compatibility-aware placement idea~\cite{cassini}. It represents the closest prior network-aware placement intuition under a comparable subset-dispatch interface by reranking GPU subsets to reduce shared-link conflicts, while omitting scheduler-level mechanisms outside the dispatch boundary.
\end{itemize}

The second group contains stronger proxy variants that isolate specific design contributions:

\begin{itemize}
    \item \textbf{Ideal-\sysname{} (\textit{Ideal-BP}):} Uses the same search procedure as \sysname{}, but with ground-truth bandwidth $B(S)$ instead of the predicted $\hat{B}(S)$, providing an upper bound on \sysname{}.
    \item \textbf{Linear Proxy (\textit{LinProxy}):} Keeps \sysname{}'s search pipeline but replaces the predictor with a linear surrogate trained on the same dataset.
    \item \textbf{Pairwise Bandwidth-aware (\textit{PairBW}):} Uses pairwise bandwidth lookup with a coarse contention penalty and a non-ML greedy selector.
\end{itemize}

Our primary metric is GBE (Eq.~(\ref{algo:metric})), which normalizes dispatcher performance across cluster states and traffic profiles. Since computing the true optimum $B^*(\mathcal{A},k,\tau)$ requires exhaustive search over $C(\mathcal{A},k)$ and is intractable in our setting, we use a best-found reference bandwidth:
\[
\tilde{B}^*(\mathcal{A},k,\tau) \triangleq \max_{S \in \mathcal{C}_{\mathrm{eval}}} B(S,\tau),
\]
where $\mathcal{C}_{\mathrm{eval}}$ is the union of candidates explored by our pruned brute-force search and all evaluated dispatchers. We compute GBE by replacing $B^*(\mathcal{A},k,\tau)$ in Eq.~(\ref{algo:metric}) with $\tilde{B}^*(\mathcal{A},k,\tau)$. This best-found normalization is applied uniformly to all dispatchers under the same evaluation candidate pool. 

\begin{figure}[t]
  \centering
  \subfloat[$R^2$ on Het-4Mix cluster\label{fig:Trained_on_H100_cluster}]{%
      \includegraphics[width=0.23\textwidth]{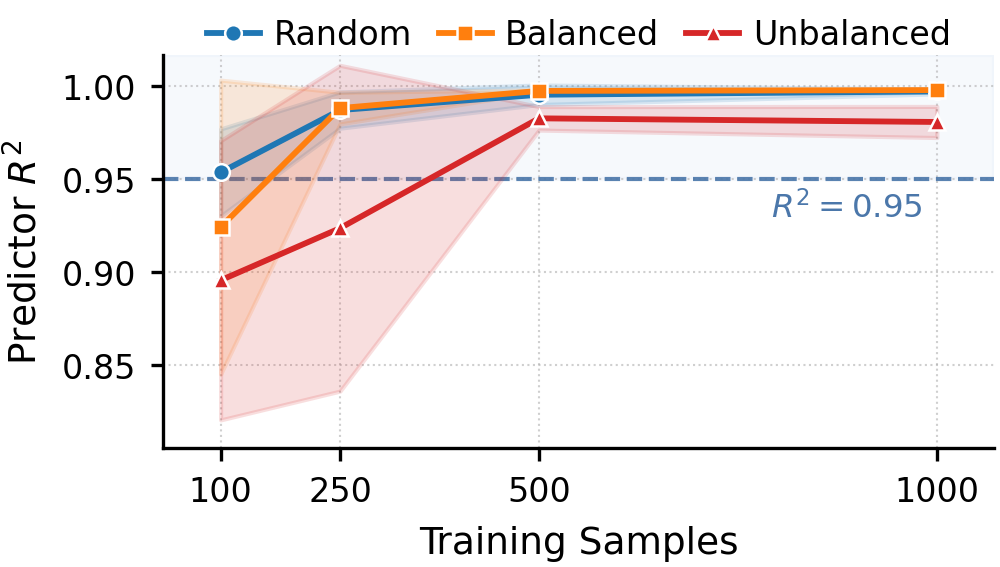}
  }
  \hfill 
  \subfloat[$MAPE$ on Het-4Mix cluster.\label{fig:Trained_on_4090V100A6000A800_cluster}]{%
      \includegraphics[width=0.23\textwidth]{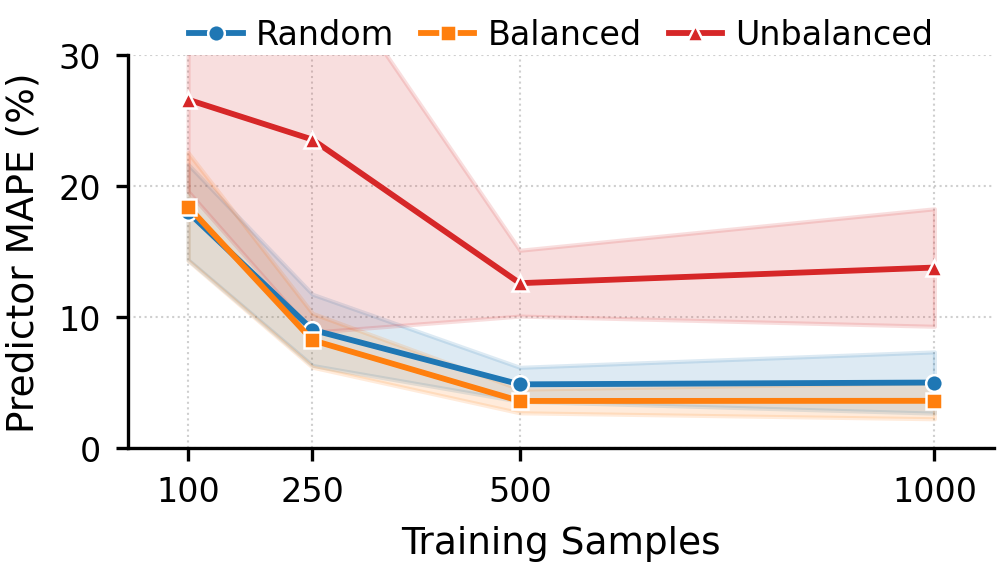}
  }
    
    \caption{Data efficiency and prediction accuracy of the Hierarchical Transformer surrogate on Het-4Mix. With randomized or balanced sampling, it reaches $R^2>0.95$ and $MAPE<10\%$ using 250 training samples.}
  \label{fig:Bandwidth Surrogate Model Accuracy}
\end{figure}


\subsection{Bandwidth Surrogate Model Accuracy}
This section evaluates the accuracy and sample efficiency of the Hierarchical Transformer for inter-host bandwidth prediction.

For the evaluation metric, the coefficient of determination ($R^2$) measures goodness-of-fit, and the Mean Absolute Percentage Error (MAPE) quantifies predictive accuracy. To rigorously test generalization, the test set is 5$\times$ larger than the training set, and all test samples are inter-host configurations, excluding trivial intra-host lookups.

Figure~\ref{fig:Bandwidth Surrogate Model Accuracy} shows that the model remains accurate under sparse profiling. Here, \emph{sparse} denotes a profiling budget that covers only a tiny fraction of the configuration space: a 32-GPU cluster has $2^{32}-1 \approx 4.3\times10^9$ non-empty GPU subsets, while our default budget uses only 250 training samples. We adopt this budget because the downstream task depends primarily on correct ranking rather than exact point estimates. Thus, a high $R^2$ is sufficient.  Larger budgets mainly reduce MAPE, while offering slight gains in ranking fidelity and additional samples can be collected after deployment.

These results show that the hierarchical structure enables the Transformer to capture the nonlinear behavior of inter-host communication from sparse measurements. 


\begin{figure*}[t]
  \centering
  \subfloat[H100, main, $\tau_{idle}$]{%
      \includegraphics[width=0.23\textwidth]{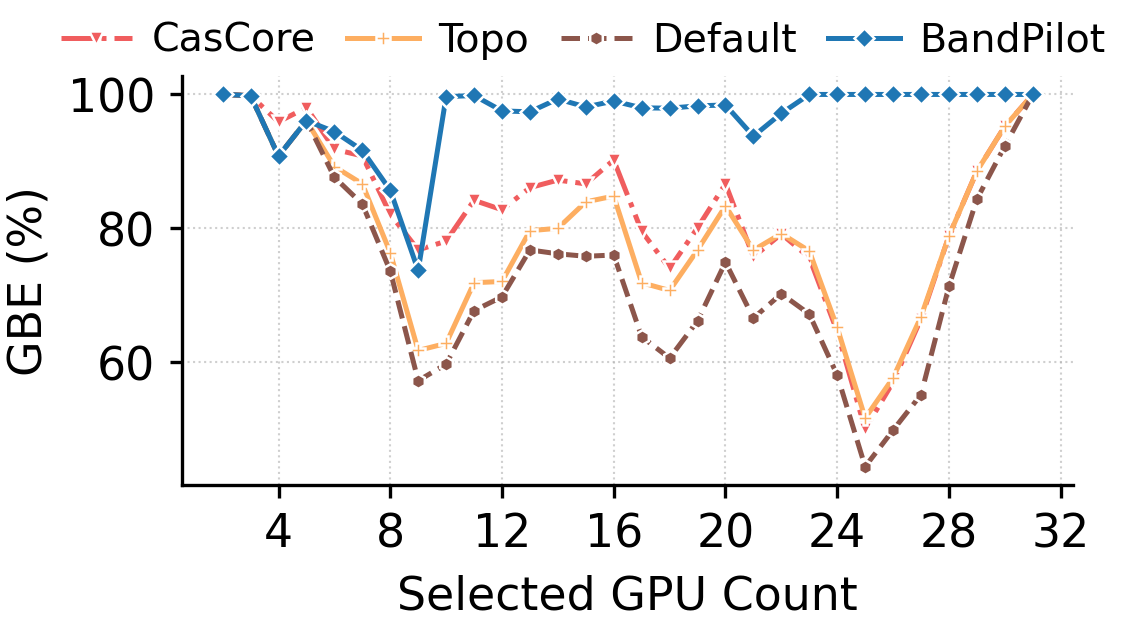}
  }
  \hfill 
  \subfloat[Het-4Mix, main, $\tau_{idle}$]{%
      \includegraphics[width=0.23\textwidth]{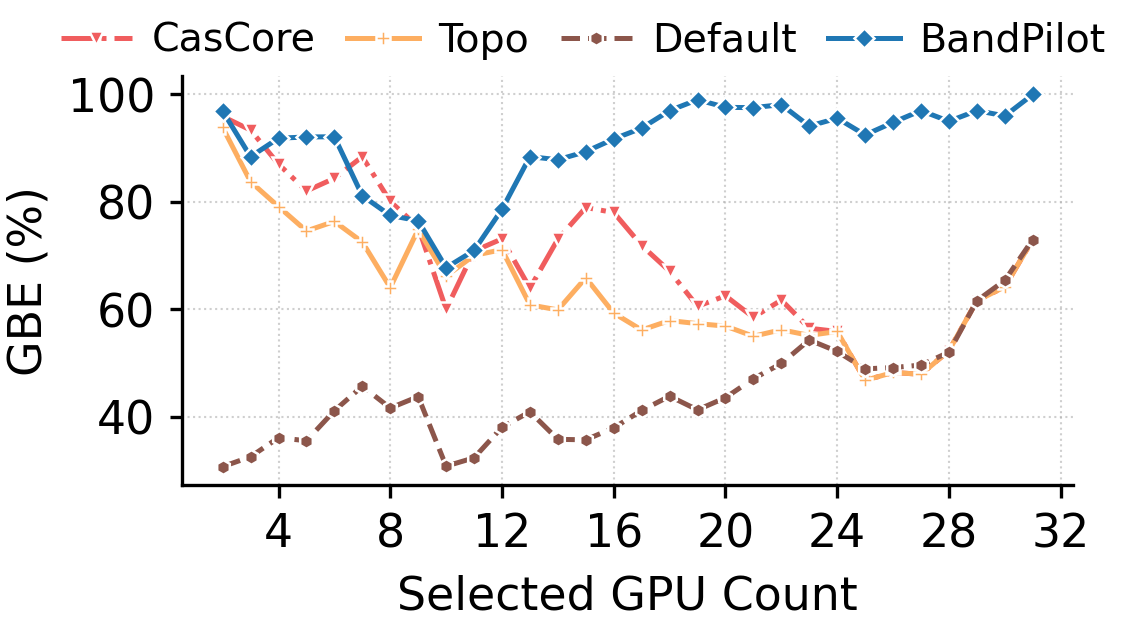}
  }
  \hfill 
  \subfloat[H100, reference, $\tau_{idle}$]{%
    \includegraphics[width=0.23\textwidth]{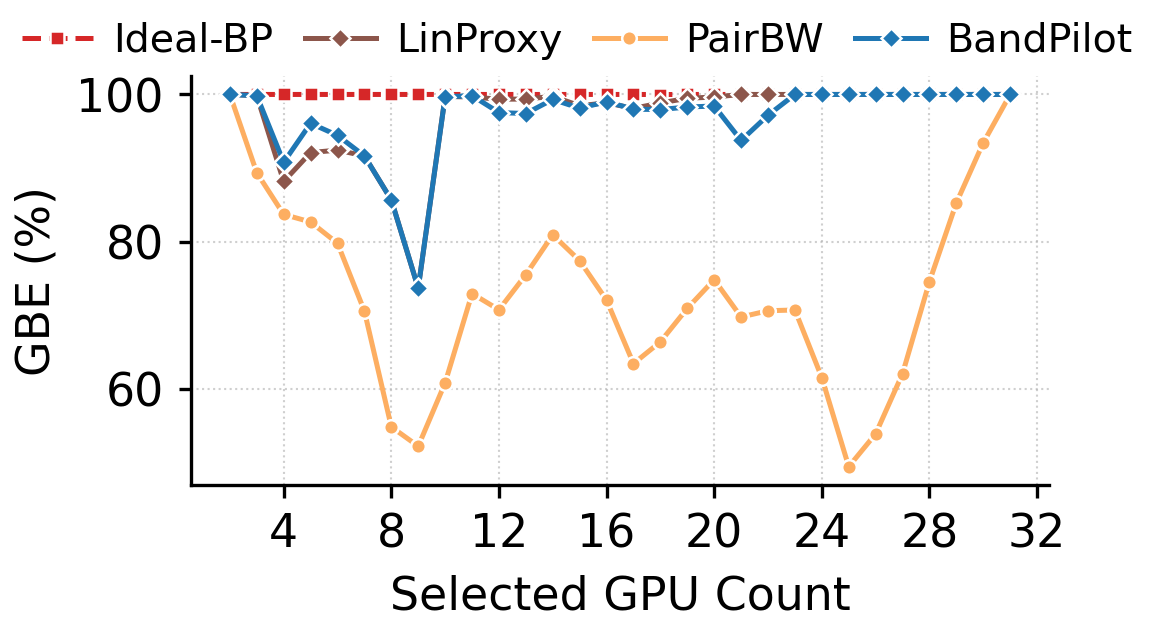}
  }
  \hfill
  \subfloat[Het-4Mix, reference, $\tau_{idle}$]{%
      \includegraphics[width=0.23\textwidth]{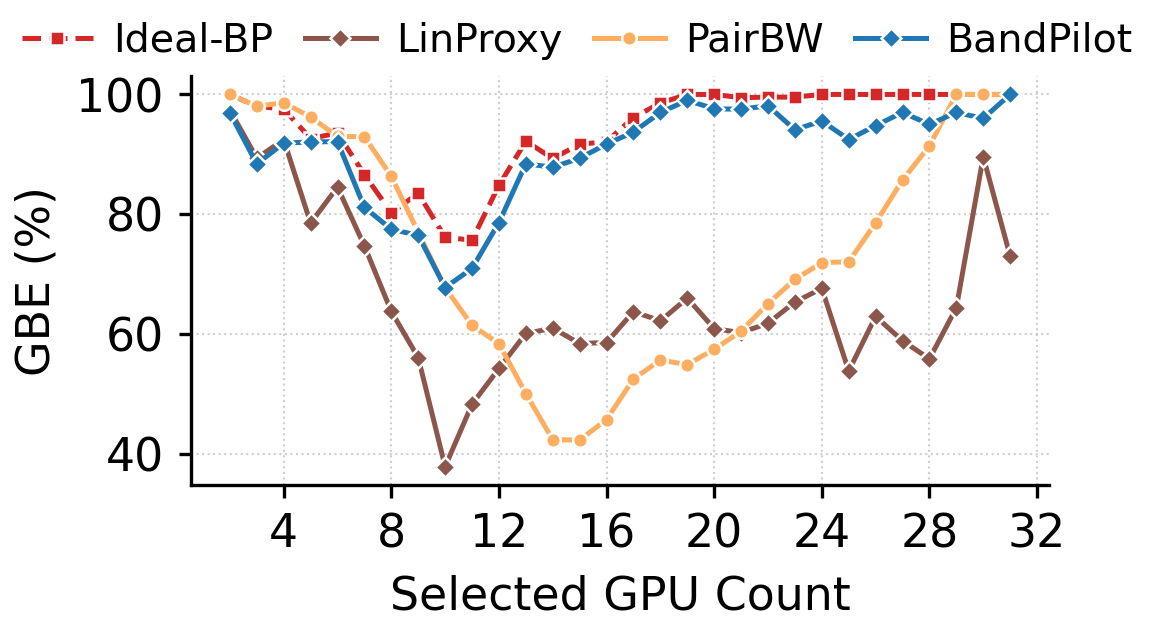}
  }
  
  \hfill 

  \subfloat[H100,main, $\tau_{mod}$]{%
      \includegraphics[width=0.23\textwidth]{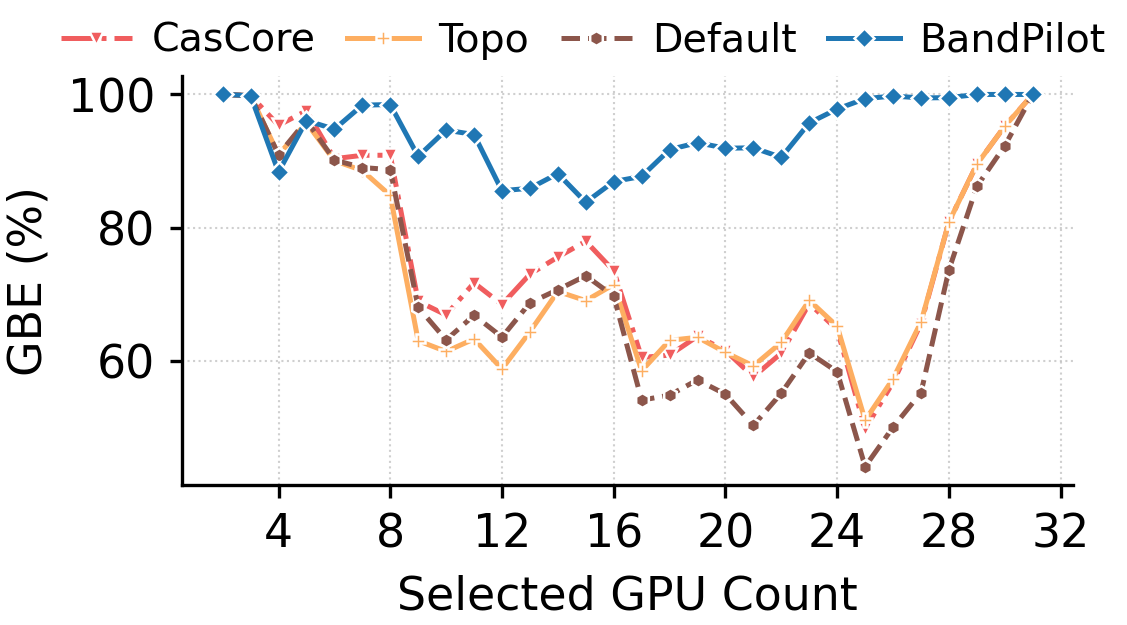}
  }
  \hfill 
  \subfloat[Het-4Mix, main, $\tau_{mod}$]{%
      \includegraphics[width=0.23\textwidth]{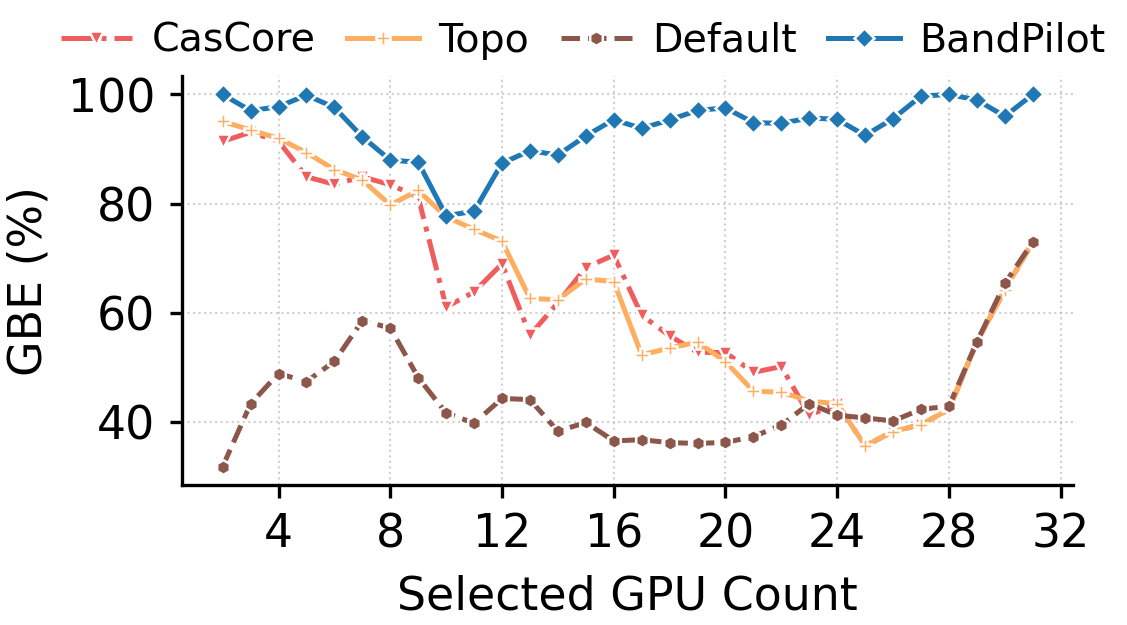}
  }
  \hfill 
  \subfloat[H100, reference, $\tau_{mod}$]{%
    \includegraphics[width=0.23\textwidth]{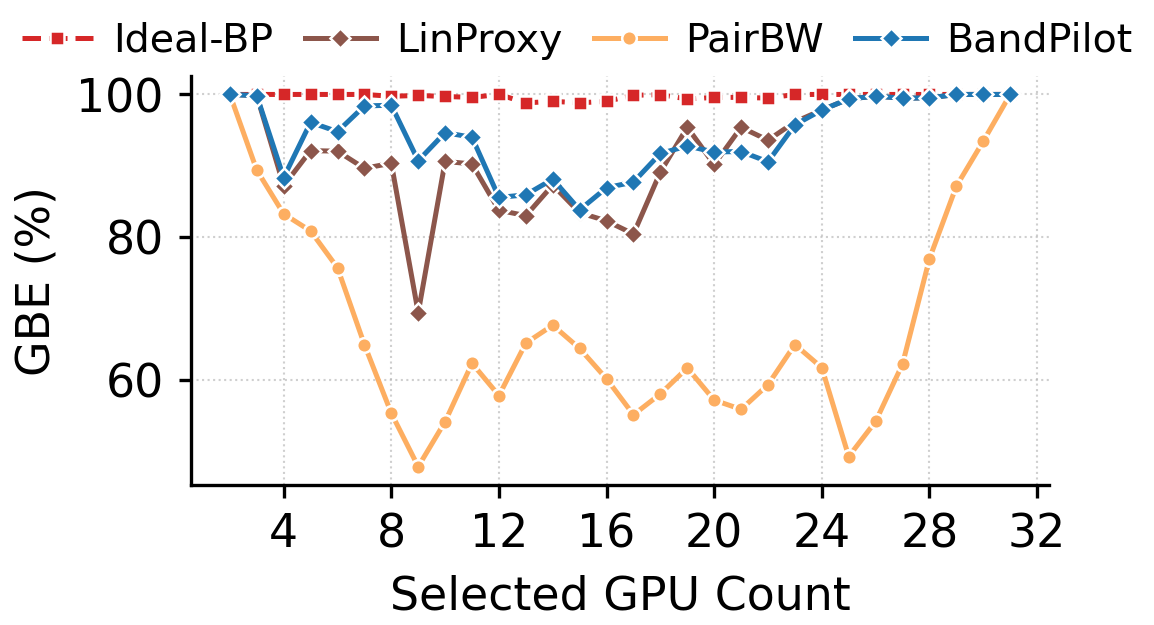}
  }
  \hfill
  \subfloat[Het-4Mix, reference, $\tau_{mod}$]{%
      \includegraphics[width=0.23\textwidth]{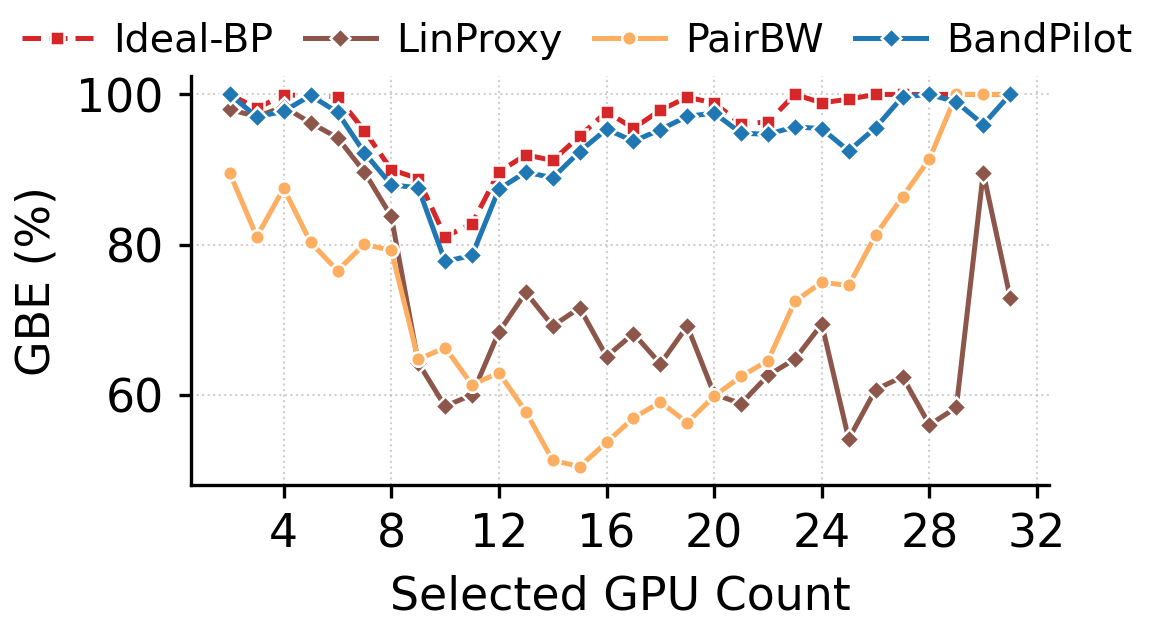}
  }

  \hfill 

  \subfloat[H100, main, $\tau_{hvy}$]{%
      \includegraphics[width=0.23\textwidth]{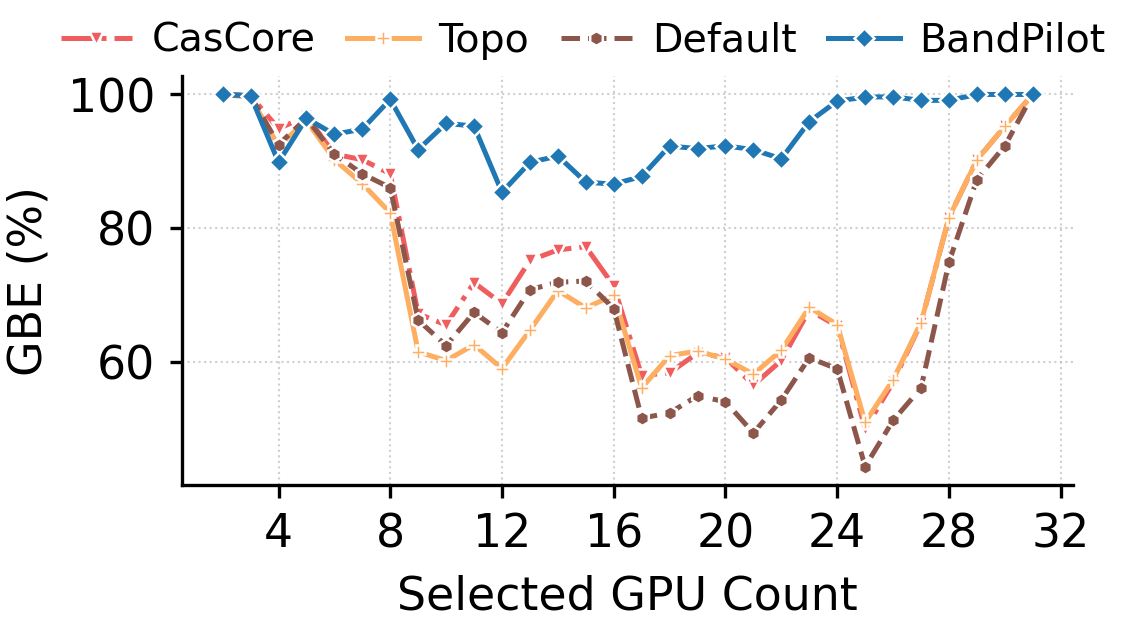}
  }
  \hfill 
  \subfloat[Het-4Mix, main, $\tau_{hvy}$\label{fig:Performance on 4090-A800 cluster under intensive}]{%
      \includegraphics[width=0.23\textwidth]{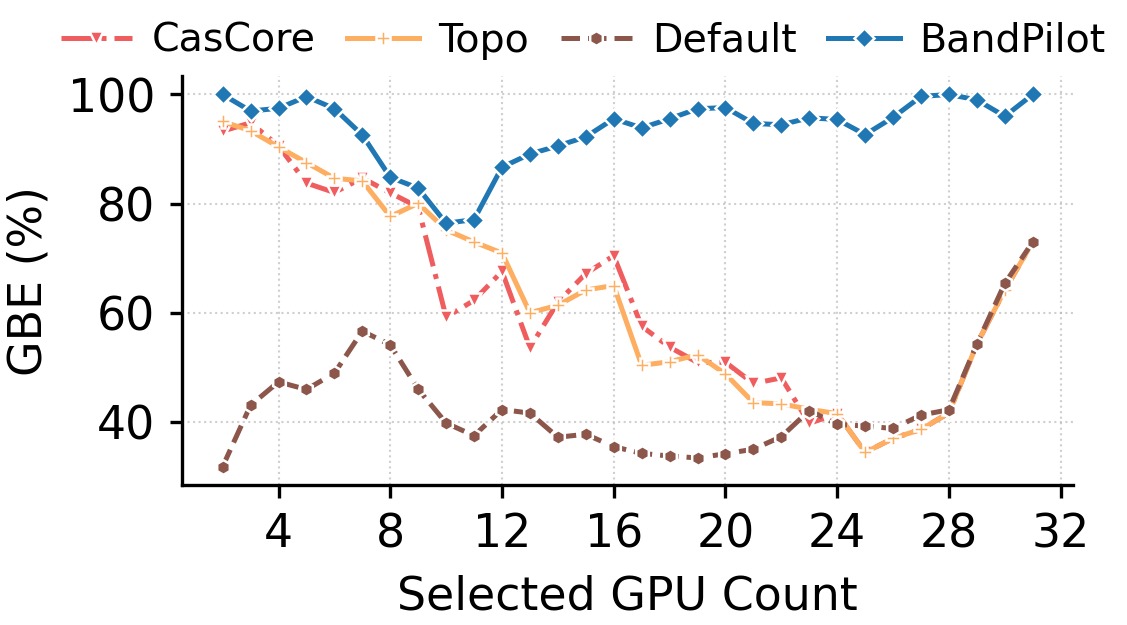}
  }
  \hfill 
  \subfloat[H100, reference, $\tau_{hvy}$]{%
    \includegraphics[width=0.23\textwidth]{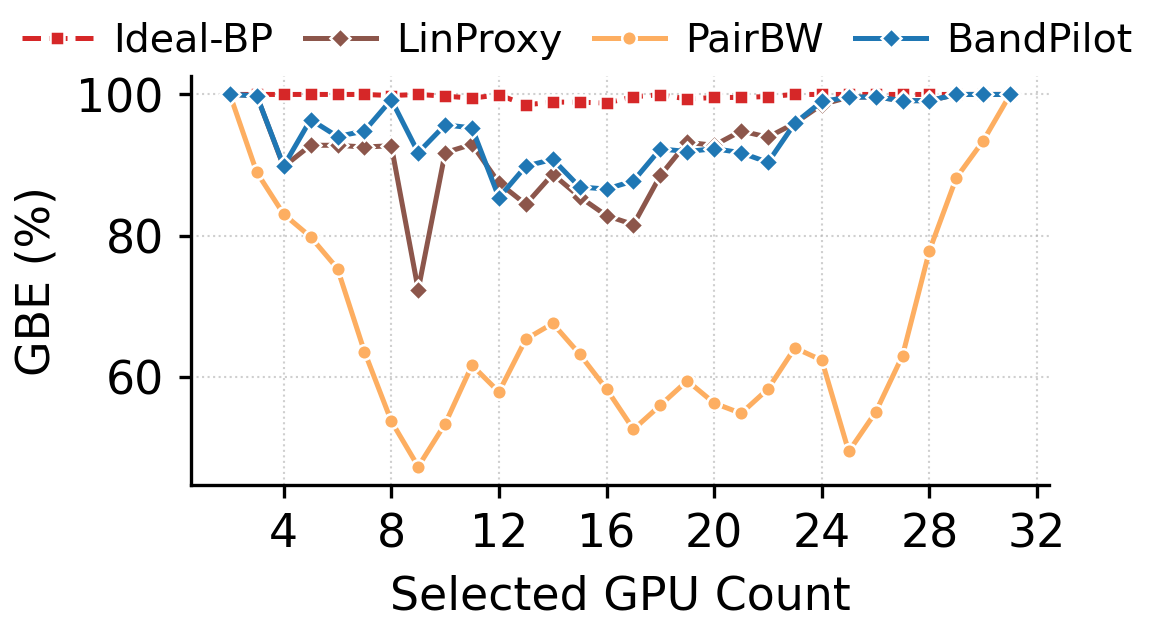}
  }
  \hfill
  \subfloat[Het-4Mix, reference, $\tau_{hvy}$]{%
      \includegraphics[width=0.23\textwidth]{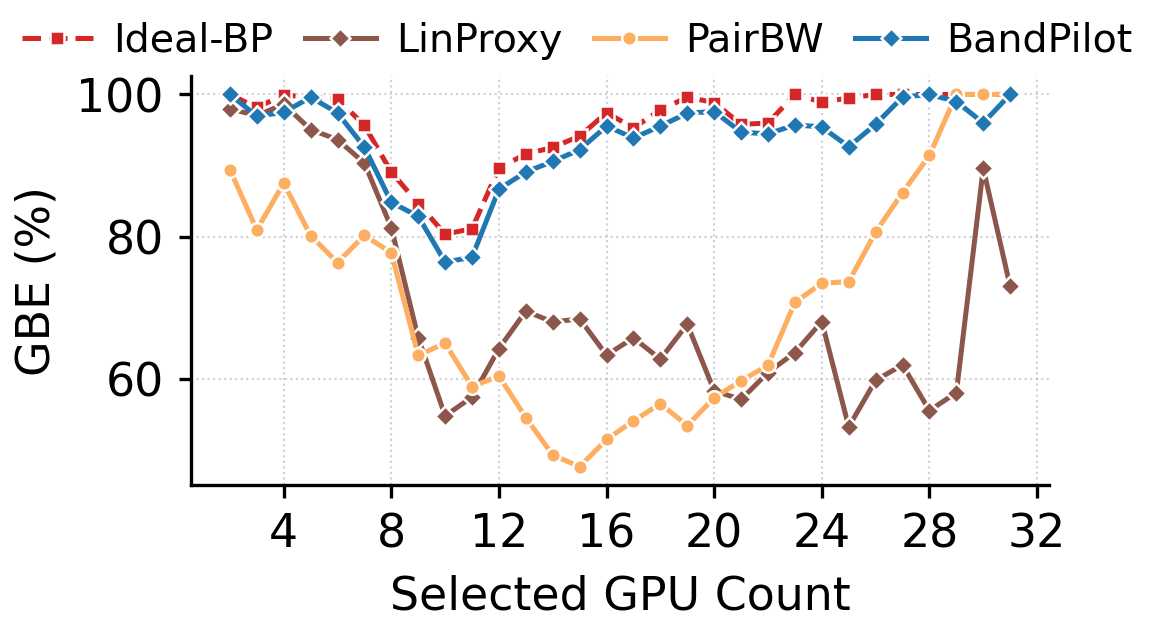}
  }
  
  \caption{Dispatching performance across different clusters under $\tau_{idle},\tau_{mod}$ and $ \tau_{hvy}$. In the main-baseline panels (a-b, e-f, i-j), \sysname{} is compared with the deployed baselines \textit{Topo} and \textit{Default}, together with the compatibility-aware \textit{CasCore}. In the reference panels (c-d, g-h, k-l), \sysname{} is further compared with oracle \textit{Ideal-BP} and the proxy baselines \textit{PairBW} and \textit{LinProxy}, which differ in the fidelity of bandwidth information available to dispatch. \sysname{} consistently outperforms the deployed heuristics and closely tracks \textit{Ideal-BP} across both clusters.
}
  \label{fig:End-to-End Dispatching Performance}
\end{figure*}


\begin{table}[htb]
  \centering
  \small
  \caption{Average Performance Comparison of Main Algorithms.}
  \label{tab:perf_comparison}
  \resizebox{\columnwidth}{!}{
  \begin{tabular}{@{}lccc@{}}
    \toprule
    \textbf{Algorithm} & \textbf{Mean GBE with $\tau_{hvy}$ (\%)} $\uparrow$ & \textbf{$\tau_{mod}$ (\%)} $\uparrow$ & \textbf{$\tau_{idle}$ (\%)} $\uparrow$  \\
    \midrule
    \addlinespace

    \multicolumn{4}{@{}l}{\textbf{H100 Cluster}} \\
    \midrule
    \textbf{\sysname{}} & 94.5 & 94.1 & 97.0 \\
    LinProxy & 92.8 & 91.9 & 97.1 \\
    CasCore & 75.4 & 75.9 & 82.7 \\
    Topo & 73.3 & 73.8 & 79.2 \\
    Default & 71.3 & 71.5 & 73.9 \\

    \addlinespace

    \multicolumn{4}{@{}l}{\textbf{Het-4Mix Cluster}} \\
    \midrule
    \textbf{\sysname{}} & 93.6 & 93.9 & 90.5 \\
    LinProxy & 70.6 & 72.0 & 66.1 \\
    CasCore & 62.2 & 63.2 & 69.1 \\
    Topo & 62.6 & 64.1 & 63.9 \\
    Default & 43.0 & 44.6 & 44.4 \\

    \bottomrule
  \end{tabular}
  }
\end{table}

\subsection{Dispatching Performance under Dynamic Cluster States}

We next evaluate whether \sysname{} delivers robust gains under dynamic multi-tenant cluster states. For each request size from 1 to 32 GPUs, we generate 50 random availability scenarios by marking a subset of GPUs as occupied, and report each dispatcher's average GBE across these scenarios. Occupied GPUs are treated as background jobs whose communication generates traffic demand according to Sec.~\ref{sec:evaluation traffic profile}. To highlight data efficiency, we train the surrogate model with 250 samples collected by balanced strategy.

Fig.~\ref{fig:End-to-End Dispatching Performance} shows two main findings. First, in the main-baseline panels, \sysname{} consistently outperforms \textit{Default}, \textit{Topo}, and \textit{CasCore} across all traffic profiles. The gap is already clear on the homogeneous H100 cluster, indicating that dispatch choice materially affects collective bandwidth even when hardware is uniform, and it becomes much larger on Het-4Mix. \textit{Topo} improves over \textit{Default} mainly on Het-4Mix but only modestly on H100, suggesting that explicit topology helps, yet static compactness remains an inadequate proxy for collective bandwidth. \textit{CasCore} further improves slightly over \textit{Default} and \textit{Topo}, but still trails \sysname{} by a wide margin, indicating that locality and conflict-avoidance alone cannot replace explicit contention-aware bandwidth modeling.


Second, in the reference panels, \sysname{} closely tracks the oracle \textit{Ideal-BP}, showing that the learned surrogate preserves the ranking information needed by dispatch. The simpler proxies reveal where this fidelity matters. On H100, \textit{LinProxy} is already competitive, suggesting that a lightweight proxy can capture much of the homogeneous case. On Het-4Mix, however, neither \textit{LinProxy} nor \textit{PairBW} matches \sysname{}, showing that richer holistic modeling becomes necessary when hardware heterogeneity and contention interact.

As summarized in Table~\ref{tab:perf_comparison}, \sysname{} achieves an average GBE of 90\%--97\% across scenarios, improving over the topology-compactness heuristic by about 20\%--30\%.

\subsection{System Overhead Analysis}

We analyze \sysname{}'s overhead in three parts: one-time profiling, online dispatch, and low-frequency adaptation.

\textbf{One-time profiling.} The one-time bring-up overhead includes: (i) intra-host profiling for each host type using \textit{nccl-tests}, which takes 10--20 min per 8-GPU host and can be executed in parallel across hosts of the same type; 
(ii) inter-host profiling across the cluster, which takes another 10--20 minutes to collect about 250 samples;  (iii) inter-host surrogate training, which takes only 1 minute; and (iv) storage overheads of 12 KB per host for the intra-host dictionaries and 354 KB for the surrogate model. The inter-host profiling data can be discarded after training.

\textbf{Online overhead.}  Fig.~\ref{fig:searching_time} shows that \sysname{} maintains practical control-plane latency. On the measured 32-GPU H100 cluster, dispatch latency stays below 0.10~s across request sizes. The gap to the shaded upper bound indicates that the KNN gate effectively prunes unnecessary refinements.  
In the large-scale Het-4Mix stress test, latency remains at the second level up to 1024 GPUs, showing that \sysname{} keeps scheduling overhead manageable at thousand-GPU scale.
To isolate the KNN gate, we further compare \sysname{} with an always-run
EHA+PTS variant over 9000 paired requests from the two clusters and three
contention modes. The KNN-gated policy reduces average dispatch latency from 54.46 ms to 34.52 ms, a 36.6\% reduction, while changing average GBE only from 95.05\% to 94.82\%, with a GBE change within a 1 percentage-point non-inferiority margin. During the same run, the gate skips PTS for 66.9\% of requests, while its decision overhead is 1.85 ms on average.



\textbf{Low-frequency model training.}
Adaptation is outside the online dispatch path and is triggered only by topology events or sustained prediction drift, as discussed in Sec.~\ref{sec:discussion}. A refresh takes about 1 minute when using measurements collected during deployment, and requires additional profiling time only when new samples must be collected.

Compared with topology-aware schedulers that require manually maintained topology descriptions such as \texttt{topology.conf}, \sysname{} replaces manual encoding with scriptable profiling and infrequent model refresh.

\begin{table}[ht]
\centering
\small
\setlength{\tabcolsep}{5pt}
\renewcommand{\arraystretch}{1.05}
\caption{Predictor ablation on Het-4Mix cluster. The hierarchical structure is markedly more sample-efficient than a naive monolithic Transformer.}
\label{tab:ablation-hier-vs-naive}

\begin{tabular*}{\columnwidth}{@{\extracolsep{\fill}}c cc cc}
  \toprule
  \multirow{2}{*}{Data} & \multicolumn{2}{c}{Hierarchical} & \multicolumn{2}{c}{Naive} \\
  \cmidrule(lr){2-3}\cmidrule(lr){4-5}
   & $R^2\uparrow$ & MAPE(\%)$\downarrow$ & $R^2\uparrow$ & MAPE(\%)$\downarrow$ \\
  \midrule
  100  & \textbf{0.97} & \textbf{19.61}  & 0.16 & 45.80 \\
  250  & \textbf{0.99} & \textbf{8.10}  & 0.22 & 45.82 \\
  1000 & \textbf{1.00} & \textbf{5.58}  & 0.44 & 42.01 \\
  \bottomrule
\end{tabular*}
\end{table}

\subsection{Ablation Study}
\label{sec:ablation study}
To validate our core design decisions, we conduct an ablation study analyzing the primary components of \sysname{}: the hierarchical architecture of the bandwidth surrogate model and the hybrid nature of the search algorithm.  


\subsubsection{Hierarchical vs. Naive strategy}
A key design choice in \sysname{} is the \emph{hierarchical} structure, which decomposes the cluster into intra-host and inter-host components and learns their interactions. To validate this choice, we compare our hierarchical predictor against a \emph{naive} Transformer that treats the full cluster topology as a monolithic input. 

Table~\ref{tab:ablation-hier-vs-naive} shows that the hierarchical design is substantially more data-efficient.
With 250 training samples, the hierarchical predictor achieves $R^2=0.99$ and 8.1\% MAPE, while the naive variant lags behind at 5$\times$ higher error.
Even with 1000 samples, the hierarchical model remains more accurate.
Overall, our hierarchical approach dramatically reduces the learning burden, leading to superior data efficiency and predictive accuracy. 

\begin{figure}[t]
  \centering
  \subfloat[Time Cost on H100 cluster]{%
      \includegraphics[width=0.23\textwidth]{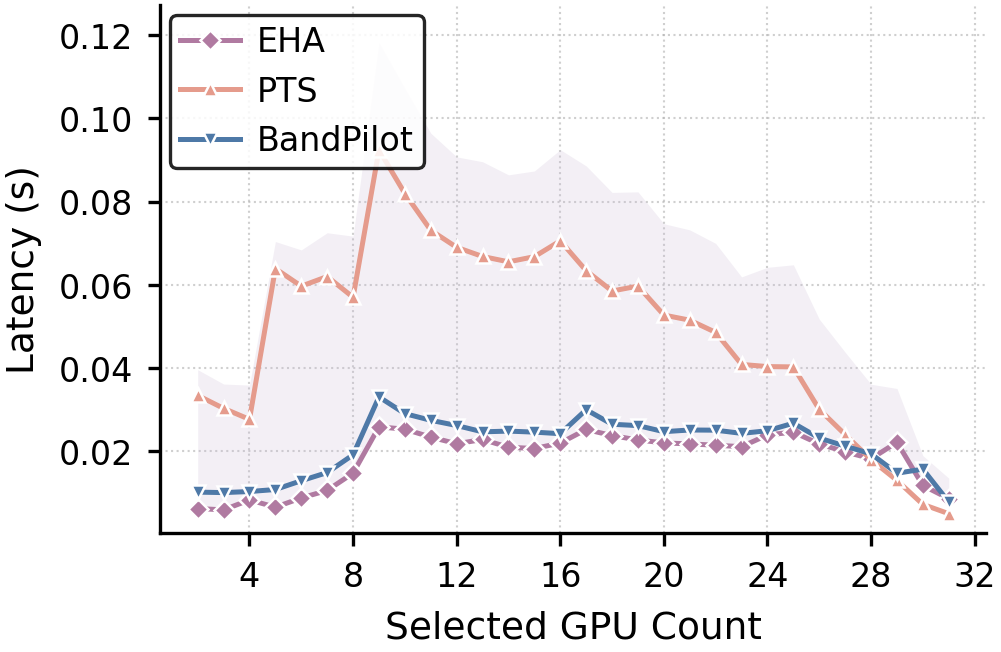}
  }%
  \hfill
  \subfloat[On large-scale Het-4Mix]{%
      \includegraphics[width=0.23\textwidth]{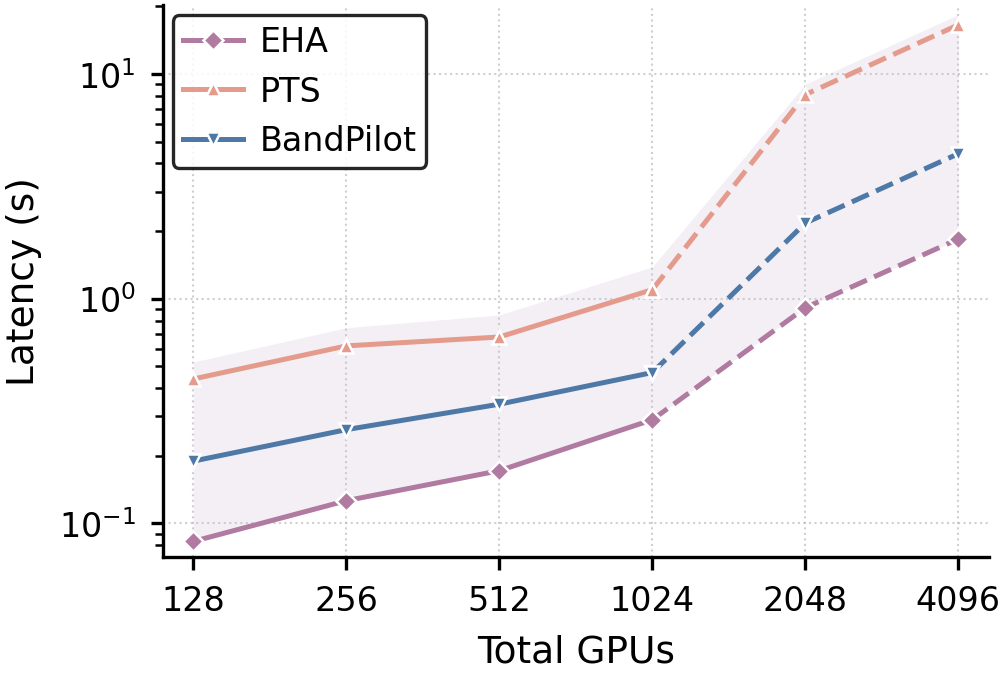}
  }
  \caption{Search-time overhead under $\tau_{mod}$. (a) Measured dispatch latency on the 32-GPU H100 cluster. The shaded region between EHA and EHA+PTS shows the overhead envelope of \sysname{} under different KNN-gate routing outcomes. (b) Trace-based Het-4Mix scale-out overhead study for 64-GPU requests ($k=64$), using the scale-out setup described in Sec.~\ref{sec:Hardware_Platforms}. This panel reports algorithmic search-time scaling as the number of equivalent scheduling units increases.}
  \label{fig:searching_time}
\end{figure}



\subsubsection{Ablation of the Hybrid Search Strategy}
We ablate the hybrid search pipeline to isolate the roles of EHA, PTS, and the KNN gate, treating EHA and PTS as standalone dispatchers and comparing them with the full \sysname{}.

\begin{figure}[t]
  \centering
  \subfloat[On H100 cluster\label{fig:Ablation on H100 cluster}]{%
      \includegraphics[width=0.23\textwidth]{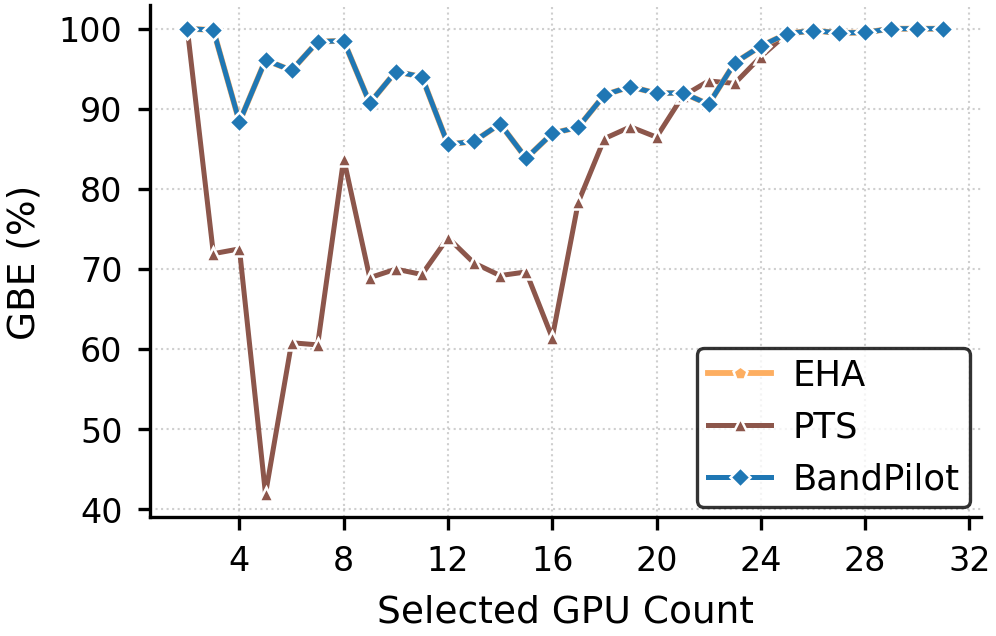}
  }
  \hfill
  \subfloat[On Het-4Mix cluster.\label{fig:Ablation on Het-4Mix cluster}]{%
      \includegraphics[width=0.23\textwidth]{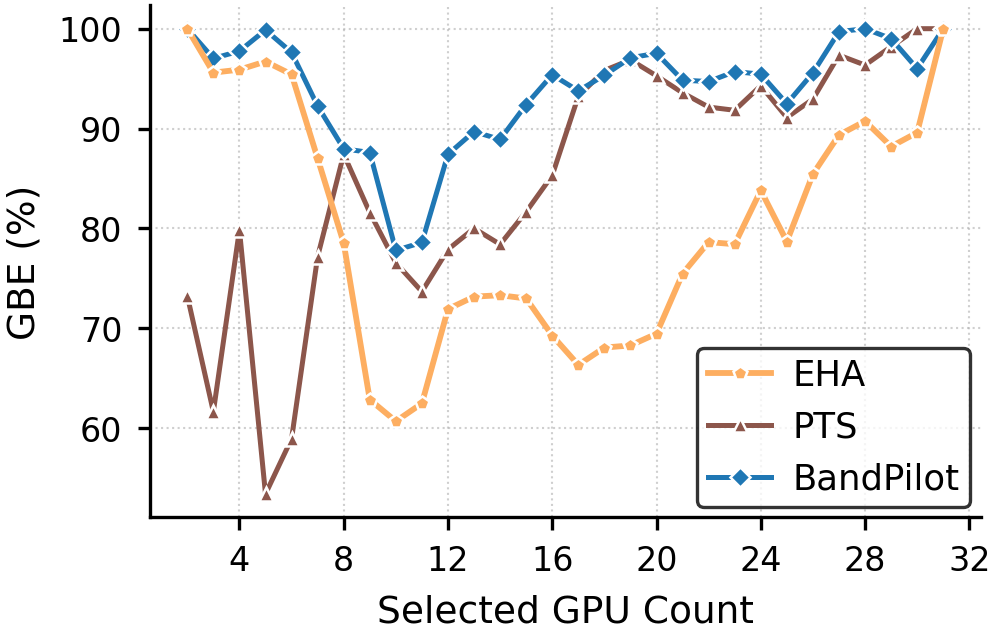}
  }

  \caption{Ablation analysis of the search pipeline on two clusters, comparing EHA, PTS, and the full \sysname{} system.}
  \label{fig:Ablation analysis of hybrid search}
\end{figure}

Fig.~\ref{fig:Ablation analysis of hybrid search} reveals a clear division of labor. On the homogeneous H100 cluster (Fig.~\ref{fig:Ablation on H100 cluster}), EHA nearly matches the full system, indicating that equilibrium-style load balance is already a strong proxy for bandwidth when the fabric is uniform. In this regime, deeper refinement is rarely necessary. On the heterogeneous Het-4Mix cluster (Fig.~\ref{fig:Ablation on Het-4Mix cluster}), however, this proxy weakens: EHA misses many high-GBE allocations, while PTS recovers much of the gap through more systematic top-down pruning. The full \sysname{} stays close to the better standalone in both settings, confirming that the hybrid design adapts to both regular and heterogeneous hardware.

Combined with Fig.~\ref{fig:searching_time}, these results motivate the KNN-gated design. \sysname{} always runs EHA first and invokes PTS only when the predicted gain justifies the extra search, retaining most of the refinement benefit without paying its cost on every request. Unless otherwise noted, all reported GBE and latency already include the KNN gate's online warm-up and update overhead.

\section{Discussion and Limitations}
\label{sec:discussion}

%

\textbf{Contention abstraction.}
\sysname{} estimates contention-degraded bandwidth with a linear proportional-sharing abstraction. This design is appropriate for a control-plane dispatcher:  it needs a stable ranking signal that captures persistent bottleneck sharing before placement, not an exact reconstruction of packet-level congestion after placement. The abstraction is most useful for collective-heavy AI workloads whose communication repeatedly stresses GPU groups and shared inter-host resources. It is less precise for highly bursty traffic whose bottleneck changes faster than the telemetry window. For localized traffic whose footprint overlaps the host, NIC, or upstream-fabric resources, the pairwise super-allocation and minimum-bottleneck construction bias the ranking away from candidates that share those resources.

%

\textbf{Model-maintenance boundary.}
\sysname{} separates online dispatch from low-frequency model maintenance.
The intra-host lookup tables remain valid as long as the host-local topology is unchanged, while the inter-host surrogate is refreshed under two classes of triggers. The first is an explicit topology event, such as introducing a new host type or changing the fabric configuration.
The second is sustained prediction drift over placements observed in production. For each completed dispatch, \sysname{} records the observed bandwidth and computes the residual against $\hat{B}(S,\tau)$. A refresh is triggered when the moving-window average residual exceeds a configurable threshold for consecutive windows. This detection relies on passive production telemetry when requests provide sufficiently fresh coverage, so \sysname{} does not place always-on active probes on the dispatch critical path. Probes are used only when passive samples are stale or sparse, after topology changes, or during maintenance-time recalibration. Fundamental fabric changes still require a new profiling round rather than lightweight fine-tuning.

\section{Related Work}

\sysname{} targets the \emph{GPU dispatch} boundary in GPU-as-a-Service clusters: given a request for \(k\) GPUs and the current cluster state, it chooses the concrete physical GPU subset before the job begins. This interface is adjacent to, but distinct from, both cluster scheduling and communication optimization. Unlike cluster schedulers, \sysname{} does not decide which job to run; unlike communication systems, it does not optimize collectives after the participant set is fixed. Instead, it provides a workload-agnostic, contention-aware dispatching primitive that can be embedded in either class of systems.

\textbf{GPU cluster schedulers and placement.}
A large body of work on multi-tenant GPU clusters optimizes utilization, fairness, or job completion time, often treating placement primarily as a locality, affinity, or packing constraint~\cite{gandiva,tiresias,antman,pollux,lyra,choi2022serving}. More recent systems make placement topology- or variability-aware, including topology-aware schedulers for multi-GPU servers and policies such as MAPA, TAMG, PAL, and Optimus~\cite{amaral2017topologygpu,ranganath2021mapa,ayadi2023tamg,jain2024pal,jiang2020unified,peng2018optimus}. Node-level studies likewise show that GPU subset choice matters under heterogeneous PCIe/NVLink connectivity~\cite{faraji2016topology}. \sysname{} is closest in spirit to this line of work, but differs in both objective and abstraction: rather than using topology as a heuristic proxy for locality, it predicts the effective collective bandwidth of each candidate subset, including cross-node effects and contention, and decouples \emph{which job to schedule} from \emph{which GPUs to assign}.

\textbf{Network- and contention-aware scheduling.}
Another line of work treats the data-center network as a first-class scheduling resource by co-optimizing compute placement, routing, or traffic sharing~\cite{whennetworkmatters,yu2022rar,han2025vclos,cassini,kumar2020mpi,lee2023optimalnetdl}. These systems reason about flow- or job-level contention, but typically assume that each job already has a fixed set of GPUs. \sysname{} is complementary: it acts \emph{before} runtime congestion management and avoids structurally poor allocations during GPU-set selection itself. In other words, \sysname{} reduces placement-induced bottlenecks upstream, while network schedulers regulate contention downstream after placement has been fixed.

\textbf{Communication optimization after device selection.}
Systems such as Crux, TACCL, AutoCCL, ACCL, Blink, SCCL, and TACOS optimize collective routing, algorithm synthesis, or library behavior for a given device group and fabric~\cite{cao2024crux,shah2023taccl,xu2025autoccl,acclplus,blink,sccl,tacos}. Their input is a fixed participant set. \sysname{} addresses the earlier question of \emph{which} GPUs should participate in the first place. The two directions are therefore orthogonal and naturally composable: \sysname{} can supply a better GPU subset, after which collective optimizers can further improve communication within that subset.

\textbf{Topology alignment and interconnect characterization.}
At larger scales, topology-aware mapping work studies how to align parallelism dimensions with hierarchical networks to reduce communication cost, again assuming that the job's GPU pool is already determined~\cite{cao2025arnold}. A related body of work benchmarks GPU interconnects and connects such measurements to collective performance~\cite{EvaluatingModernGPUInterconnect,hidayetoglu2024commbench}. \sysname{} leverages the same measurement interface to train its performance surrogate, supporting contention-aware dispatch in a live multi-tenant cluster.

\section{Conclusion}
We presented \sysname{}, a performance- and contention-aware GPU dispatching system that replaces static topology heuristics with data-driven performance prediction. By estimating the communication bandwidth of candidate allocations from sparse measurements, \sysname{} efficiently identifies near-optimal dispatch decisions under multi-tenant interference. Across a physical H100 cluster and heterogeneous simulations, \sysname{} consistently achieves high bandwidth efficiency and outperforms SOTA dispatchers. This work demonstrates that moving beyond static proxies to performance- and contention-aware dispatching is viable and important for unlocking the potential of GPU clusters.

\bibliography{Reference}
\bibliographystyle{IEEEtran}

\begin{IEEEbiography}[{\includegraphics[width=1in,height=1.25in,clip,keepaspectratio]{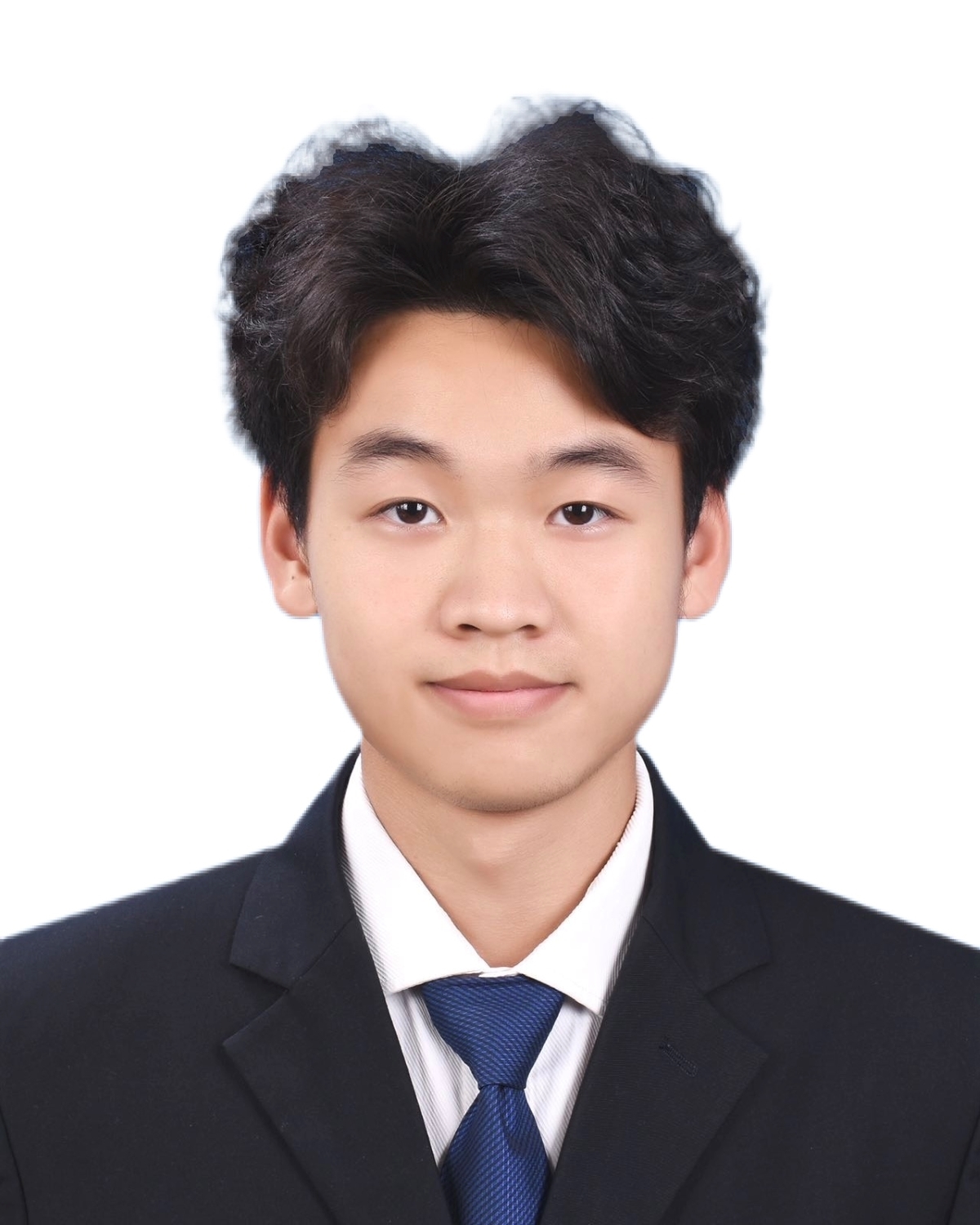}}]{Kunming Zhang} is currently pursuing the Ph.D. degree in Data Science and Analytics at The Hong Kong University of Science and Technology (Guangzhou). He received the bachelor's degree in Management from Xi'an Jiaotong University. His research interests include green computing, cloud computing, and machine learning systems.
\end{IEEEbiography}

\begin{IEEEbiography}[{\includegraphics[width=1in,height=1.25in,clip,keepaspectratio]{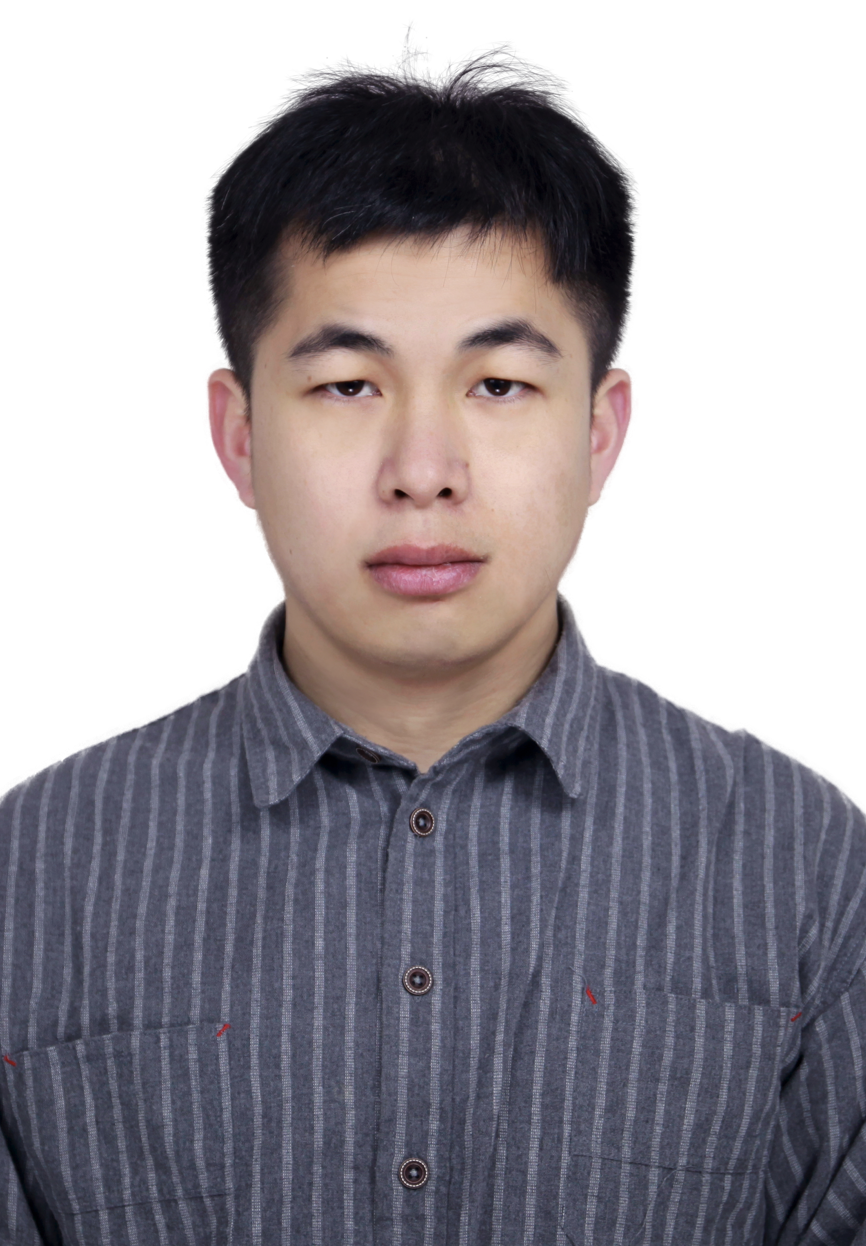}}]{Hanlong Liao} is currently pursuing the Ph.D. degree in Control Science and Engineering at the National University of Defense Technology. He received his B.S. degree in Computer Science from Beijing Institute of Technology. His research interests include LLM system optimization, cloud computing, green computing, and operations research.
\end{IEEEbiography}

\begin{IEEEbiography}[{\includegraphics[width=1in,height=1.25in,clip,keepaspectratio]{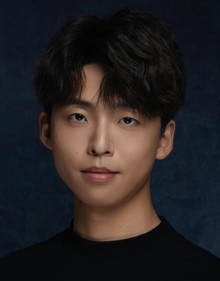}}]{Junyu Xue} is currently pursuing the Ph.D. degree in electronic information engineering at the Southern University of Science and Technology, Shenzhen, China. He received his B.Eng. degree from China University of Petroleum (East China). His research interests include machine learning systems, networked systems, and applied AI.
\end{IEEEbiography}

\begin{IEEEbiography}[{\includegraphics[width=1in,height=1.25in,clip,keepaspectratio]{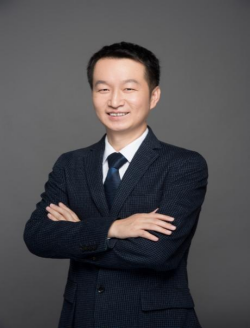}}]{Deke Guo} received the B.S. degree in industry engineering from Beijing University of Aeronautics and Astronautics, Beijing, China, in 2001, and the Ph.D. degree in management science and engineering from the National University of Defense Technology, Changsha, China, in 2008. He is currently a Professor with the School of Computer Science and Engineering, Sun Yat-sen University. His research interests include distributed systems, computer network, and LLM optimization.
\end{IEEEbiography}

\begin{IEEEbiography}[{\includegraphics[width=1in,height=1.25in,clip,keepaspectratio]{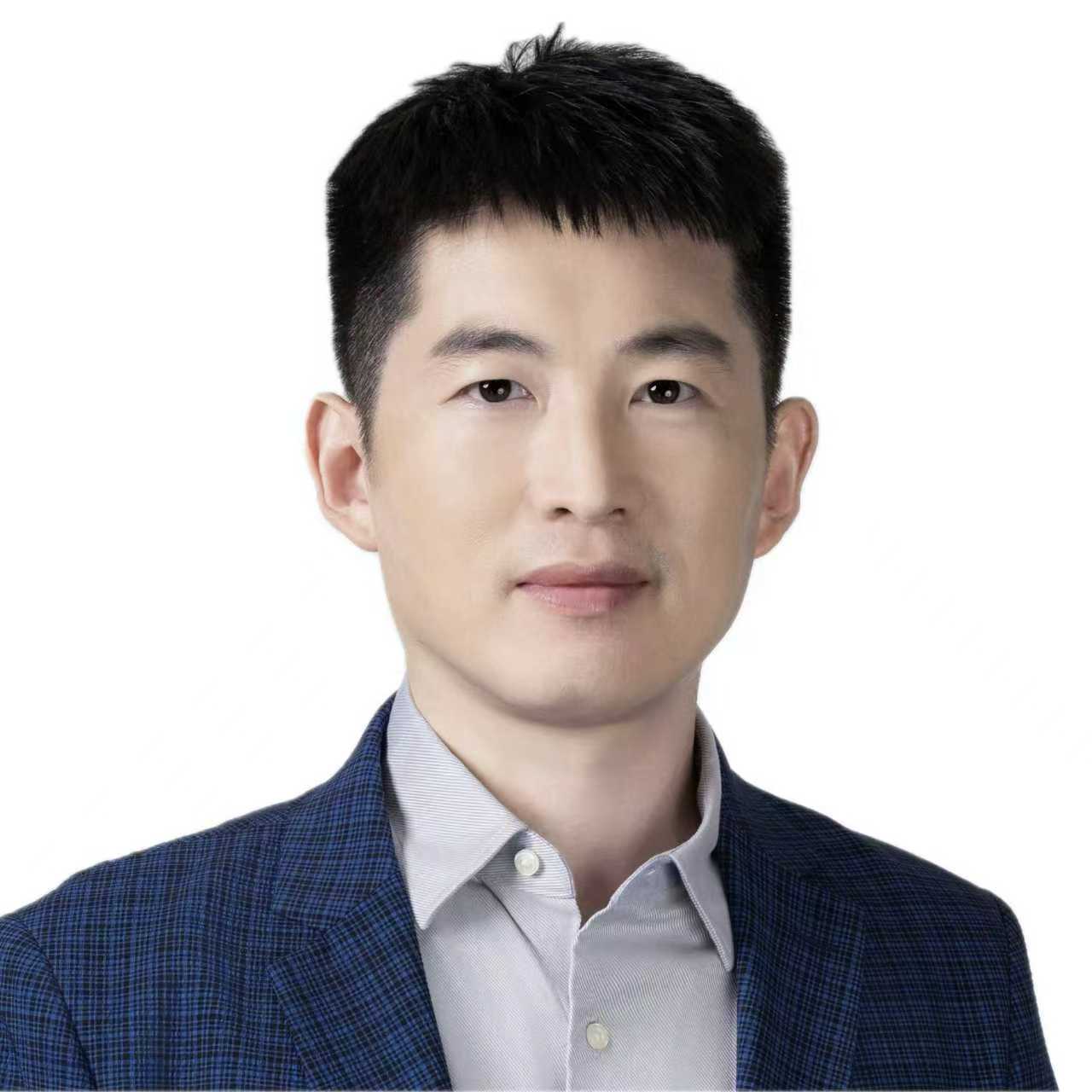}}]{Guoming Tang} is currently an Assistant Professor in the Data Science and Analytics (DSA) Thrust at the Information Hub, The Hong Kong University of Science and Technology (Guangzhou). He received his Ph.D. in Computer Science from the University of Victoria, Canada, and both his Bachelor's and Master's degrees from National University of Defense Technology, China. His research mainly focuses on green/sustainable computing, cloud/edge computing, and applied AI/ML.
\end{IEEEbiography}


\end{document}